\documentclass{emulateapj}
\usepackage{apjfonts}
\usepackage{natbib}

\def\ltsim{\raise 2pt \hbox {$<$} \kern-1.1em \lower 4pt \hbox {$\sim$}}
\def\gtsim{\raise 2pt \hbox {$>$} \kern-1.1em \lower 4pt \hbox {$\sim$}}

\begin{document}

\title[Cavities and shocks in the compact group of
galaxies HCG 62]{Cavities and shocks in the galaxy group 
   HCG 62 as revealed by {\it Chandra}, {\it XMM}
  and {\it GMRT} data}

\author{Myriam Gitti$^{1,2}$, Ewan O'Sullivan$^{1,3}$,
Simona Giacintucci$^{1,4}$, Laurence P. David$^1$, Jan Vrtilek$^1$,
Somak Raychaudhury$^3$, and Paul E. J. Nulsen$^1$}

\address{$^1$ Harvard-Smithsonian Center for Astrophysics,
60 Garden Street, Cambridge, MA 02138 - USA}
\address{$^2$ Astronomical Observatory of Bologna - INAF,
via Ranzani 1, I-40127 Bologna  - Italy}
\address{$^3$ University of Birmingham,
Edgbaston, Birmingham B15 2TT - UK}
\address{$^4$ Institute of Radioastronomy - INAF,
via Gobetti 101, I-40129 Bologna  - Italy}

\begin{abstract}

  We report on the results of an analysis of {\it Chandra}, {\it
    XMM--Newton} and new {\it GMRT} data of the X-ray bright compact
  group of galaxies HCG 62, which is one of the few groups known to
  possess clear, small X-ray cavities in the inner regions.  This is
  part of an ongoing X-ray/low-frequency radio study of 18 groups,
  initially chosen for the availability of good-quality X-ray data and
  evidence for AGN/hot gas interaction. At higher frequency (1.4 GHz),
  the HCG 62 cavity system shows minimal if any radio emission, but
  the new {\it GMRT} observations at 235 MHz and 610 MHz clearly
  detect extended low-frequency emission from radio lobes
  corresponding to the cavities.  By means of the synergy of X-ray and
  low-frequency radio observations, we compare and discuss the
  morphology, luminosity and pressure of the gas and of the radio
  source. We find that the radio source is radiatively inefficient,
  with a ratio of radio luminosity to mechanical cavity power of $\sim
  10^{-4}$, and that the radio pressure of the lobes is about one
  order of magnitude lower than the X-ray pressure of the surrounding
  thermal gas. Thanks to the high spatial resolution of the {\it
    Chandra} surface brightness and temperature profiles, we also
  identify a shock front located at 36 kpc to the south-west of the
  group center, close to the southern radio lobe, with a Mach number
  $\sim 1.5$ and a total power which is about one order of magnitude
  higher than the cavity power.  Such a shock may have heated the gas
  in the southern region, as indicated by the temperature map.  The
  shock may also explain the arc-like region of enriched gas seen in
  the iron abundance map, as this may be produced by a non-Maxwellian
  electron distribution near its front.

\end{abstract}


\maketitle


\section{Introduction}

Many possibilities have been proposed in the last decade to solve the
so-called ``cooling flow problem'' in the hot atmospheres of galaxy
clusters and groups., i.e., the lack of evidence for central gas
cooling to very low temperatures at the rates predicted (see, e.g.,
the review by Peterson \& Fabian 2006).  Among these, feedback by the
central active galactic nucleus (AGN) appears to be the most promising
solution.  Identifying radio galaxies (galaxies whose AGN produce the
radio-emitting jets) as a primary source of feedback in clusters has
been one of the major achievements of the current generation of X-ray
observatories. Although gas does cool at least through part of the
X-ray emitting temperature band ($10^7$-$10^8$ K) in galaxy clusters,
{\it Chandra} and {\it XMM-Newton} have shown that there is not a
significant amount of gas cooling below about one third of its
original temperature (e.g., Peterson et al. 2003; Kaastra et
al. 2004). These observations also reveal highly disturbed structures
in the cores of many clusters, including shocks, cavities and sharp
density discontinuities. At radio wavelengths, it is clear that AGN
jets are the cause of many of these disturbances.  The most typical
configuration is for jets from the central dominant elliptical of a
cluster to extend outwards and inflate lobes of radio-emitting
plasma. These lobes push aside the X-ray emitting gas of the cluster
atmosphere, thus leaving apparent cavities in the X-ray images (e.g.,
for a review: McNamara \& Nulsen 2007 and references therein).

There are already several in-depth analyses of individual rich
clusters (e.g., Perseus: Fabian et al. 2006; Centaurus: Sanders et
al. 2008; MS0735.6+7421: Gitti et al. 2007; A2052: Blanton et
al. 2009) and studies of cluster samples (B\^irzan et al. 2004, 2008;
Dunn et al. 2005; Dunn \& Fabian 2006; Rafferty et al. 2006; Diehl et
al. 2008). However, the differences between groups and clusters imply
that the existing studies on feedback tell us little about how it
operates in groups.  Furthermore, due to the shallower group
potential, the AGN outburst is expected to have a large impact on the
intragroup medium.  It is therefore essential to study individual
groups and group samples in order to understand how feedback has
influenced the thermal history of galaxies and the intragroup medium,
and thus of most of the baryons in the Universe.  Such investigations
have been undertaken only recently (e.g., Johnson et al. 2009; Sun
2009; Gastaldello 2008, 2009; Giacintucci et al. in prep.; O'Sullivan
et al. in prep.).

The work presented here is part of an ongoing project aimed at
combining X-ray and low-frequency radio observations of galaxy
groups. In particular, we have selected a sample of 18 galaxy groups
based on the presence of signs of interaction between the hot gas and
the central AGN. For these groups, which all have good quality X-ray
data in the archives of {\it Chandra} and/or {\it XMM--Newton}, we
have obtained new radio data at the Giant Metrewave Radio Telescope
({\it GMRT}) at 235 MHz, 327 MHz, and 610 MHz (Giacintucci et
al. 2008, 2009; Raychaudhury et al. 2009, Giacintucci et al. in
prep.).

HCG 62 (z=0.0137) is the X-ray brightest of the Hickson compact groups
and has been extensively studied with {\it ROSAT}, {\it ASCA}, {\it
  Chandra}, {\it XMM--Newton} and {\it Suzaku} (Ponman \& Bertram
1993; Finoguenov \& Ponman 1999; Buote 2000; Morita et al. 2006; Gu et
al. 2007; Tokoi et al. 2008; Sanders et al. 2009).  The central region
of this group is dominated by four early-type galaxies, two of which
are likely interacting as indicated by optical studies (Valluri \&
Anupama 1996; Spavone et al. 2006).  This was the first galaxy group
with a clear detection of inner cavities (Vrtilek et al. 2002).  The
existing 1.4 GHz VLA observations mainly show the emission from the
compact radio source and indicate only some hints of extended radio
emission toward the southern cavity (Vrtilek et al. 2002).  Owing to
the poor radio images then available, the HCG 62 cavity system was
classified as a ``radio ghost'' in the sample of B\^irzan et
al. (2004).

We present here new low frequency {\it GMRT} radio observations of HCG
62 that, together with the existing X-ray {\it Chandra} and {\it
  XMM--Newton} observations, allow us to study the X-ray/radio
interaction and investigate the AGN feedback in this group.  This
paper is organized as follows. In \S \ref{data.sec} we describe the
data sets used for our investigation and the data reduction process,
in \S \ref{disturbed.sec} we show the X-ray and radio morphologies of
the group inner region, and in \S \ref{xray.sec} we present the X-ray
properties derived from the spectral analysis. In \S
\ref{interaction.sec} we investigate the interaction of the radio
plasma with the hot gas in terms of energetics (\S
\ref{energetics.sec}) and pressure balance (\S \ref{pressure.sec}). In
\S \ref{front.sec} we analyze the surface brightness discontinuities,
finding evidence for the detection of weak shocks. Finally, in \S
\ref{maps.sec} we present the 2-D distribution of temperature (\S
\ref{tmap.sec}) and iron abundance (\S \ref{Zmap.sec}).  The summary
of our main results is given in \S \ref{summary.sec}.

With $H_0 = 70 \mbox{ km s}^{-1} \mbox{ Mpc}^{-1}$, and
$\Omega_M=1-\Omega_{\Lambda}=0.3$, the luminosity distance to HCG 62
is 59 Mpc and 1 arcsec corresponds to 0.28 kpc in the rest frame of
the group.  The radio spectral index $\alpha$ is defined such as
$S_{\nu} \propto \nu^{- \alpha}$, where $S_{\nu}$ is the flux density
at the frequency $\nu$.


\begin{figure*}[ht]
\centerline{
\includegraphics[width=8.5cm]{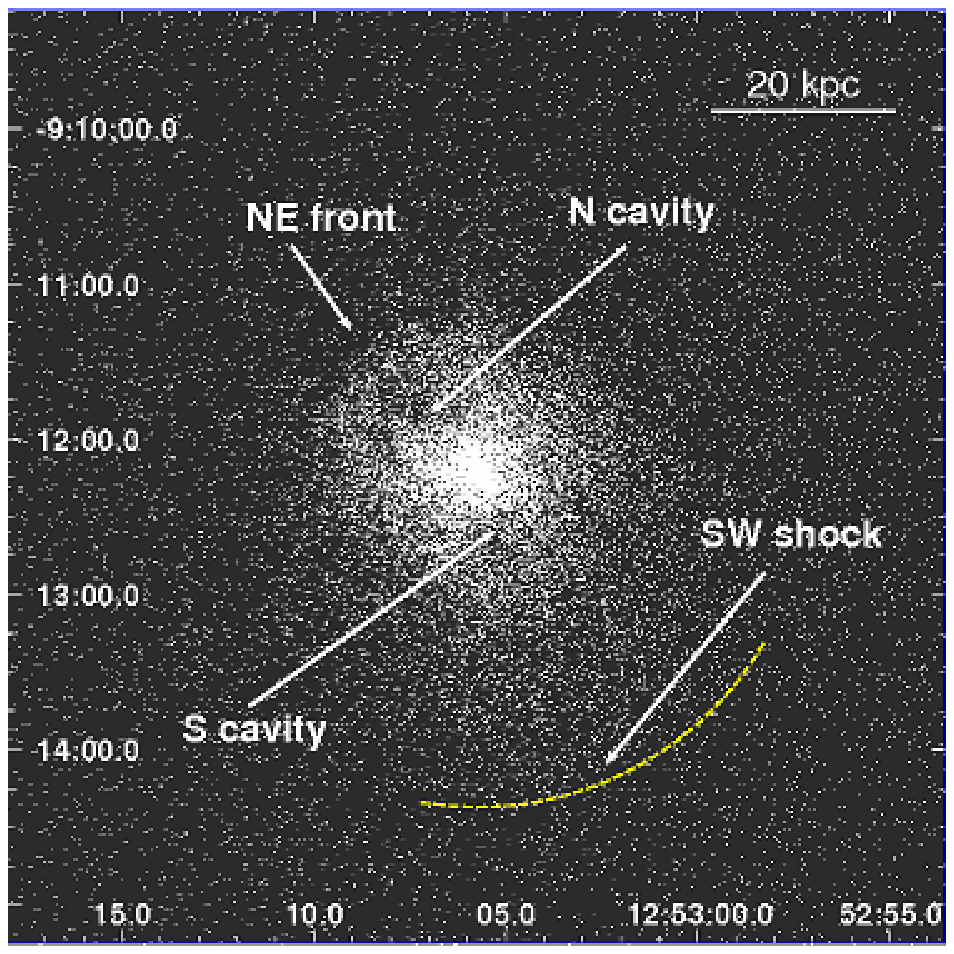}
\includegraphics[width=8.5cm]{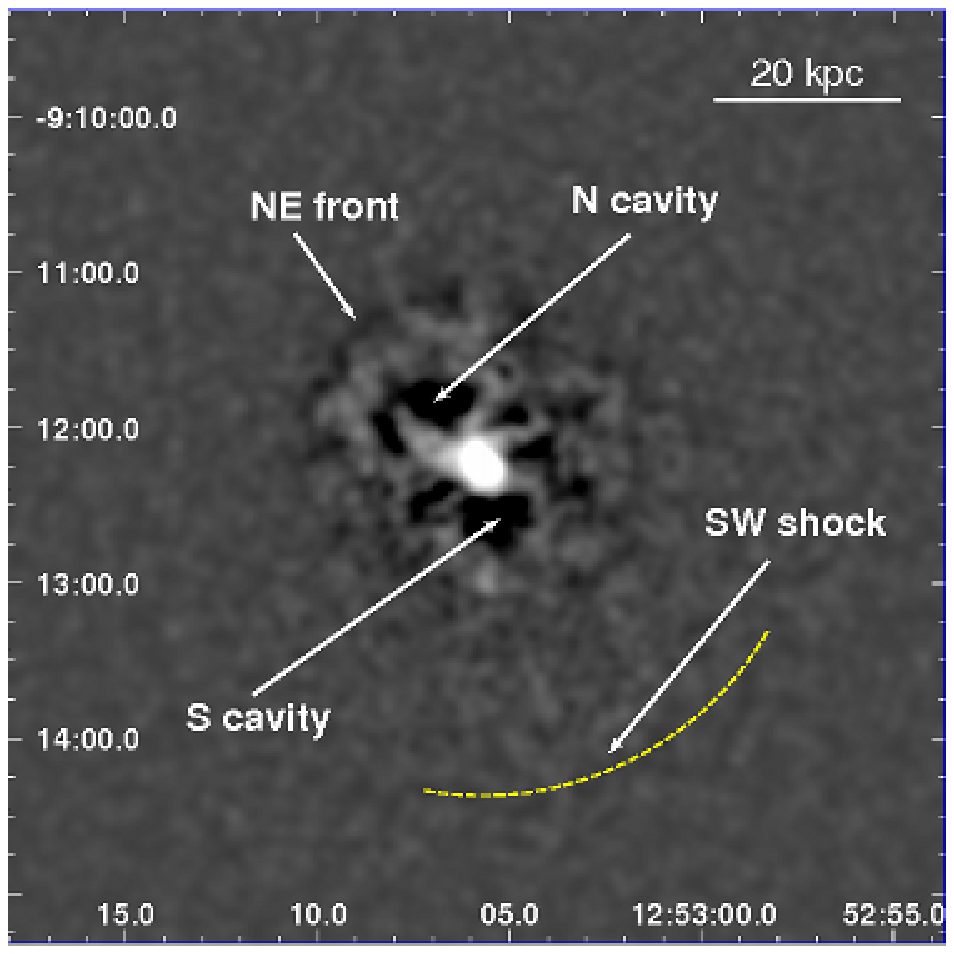}
}
\centerline{
\includegraphics[width=8.5cm]{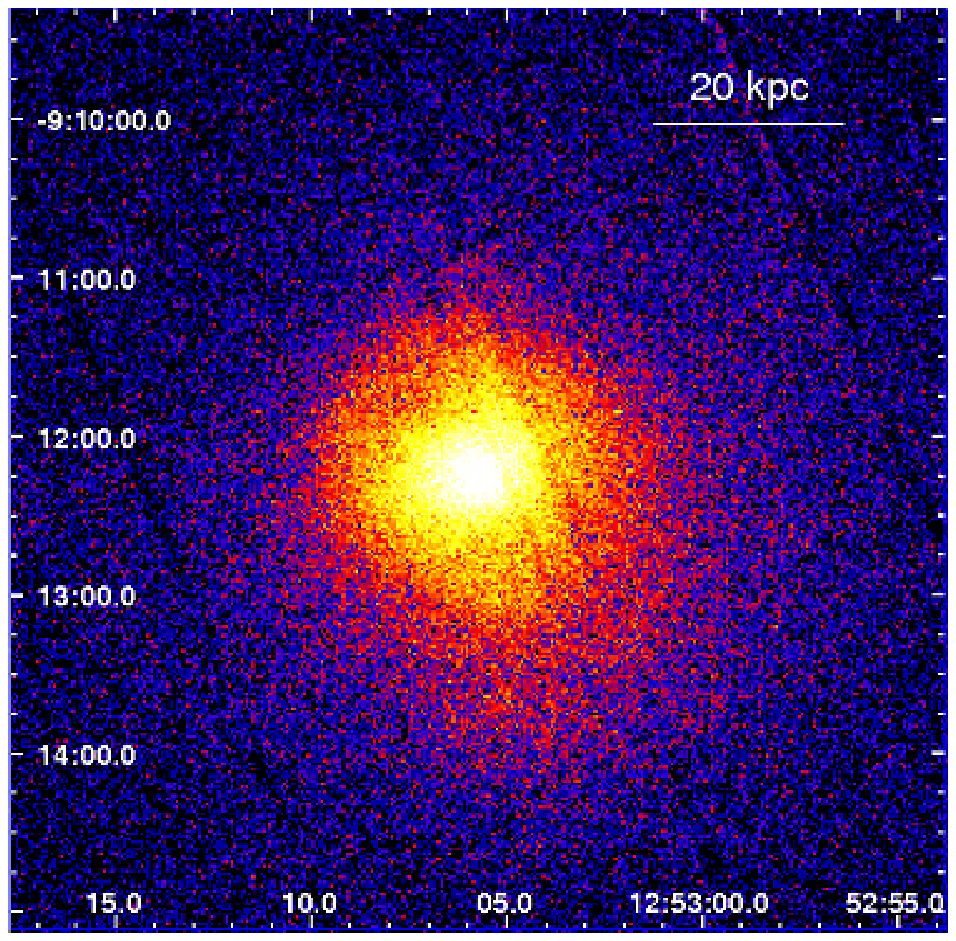}
\includegraphics[width=8.5cm]{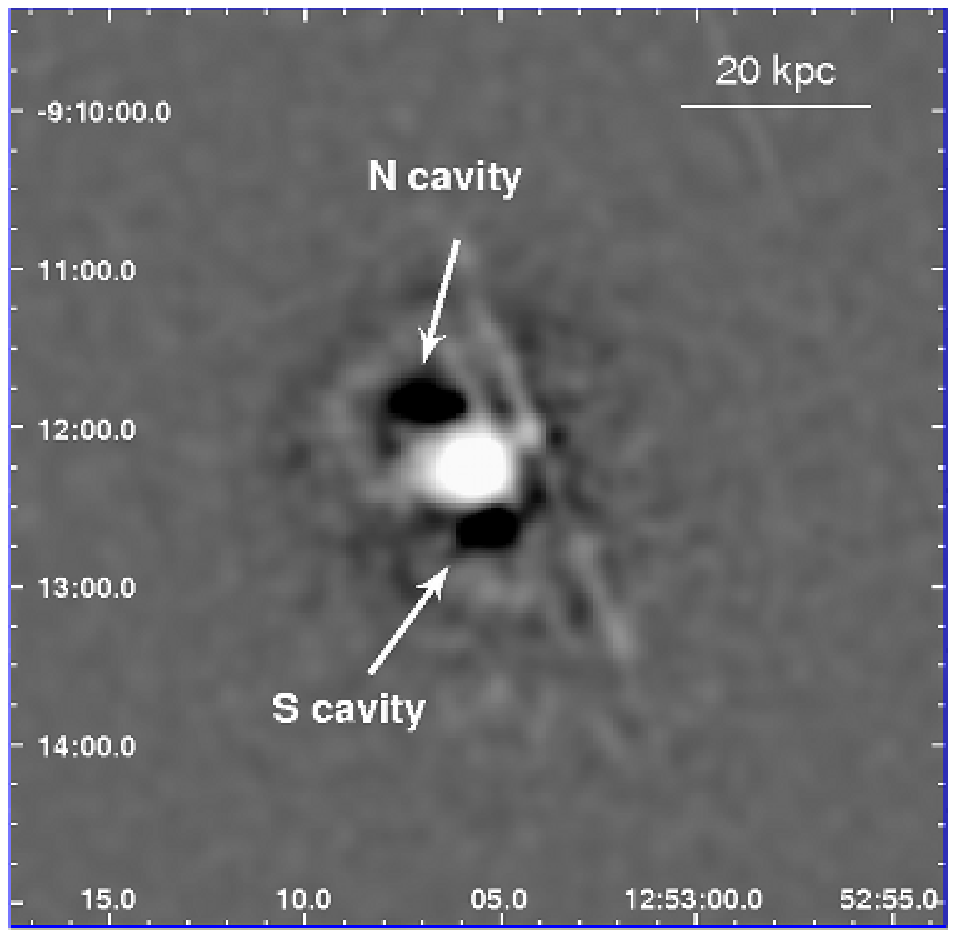}
}

\caption{\label{xmm-chandra.fig} \textit{Top:} Raw (left) and unsharp
  masked (right) 0.5-2.0 keV ACIS-S image of the central region of HCG
  62.  \textit{Bottom:} Raw (left) and unsharp masked (right) 0.5-2.0
  keV MOS1+MOS2+PN mosaic image of the central region of HCG 62.  The
  unsharp masked images are produced by subtracting a large-scale (20
  pixels) smoothed image from a small-scale (5 pixels) smoothed image.
  The arrows indicate the features discussed in the text (see \S
  \ref{X-morphology.sec}). In particular, the dashed arc in top panels
  indicates the position of the SW shock, which is non visible clearly
  in the images. In all panels, the box is $6' \times 6'$ (101 kpc by
  101 kpc). North is up and east is left. }
\end{figure*}

\section{Observations and data reduction}
\label{data.sec}

\subsection{{\it Chandra}}

HCG 62 was observed with {\it Chandra} for $\sim 48.5$ ks on 2000
January 25 (ObsID 921) with ACIS-S in imaging mode operating at the
focal plane temperature of $-110^{\circ}$C.  The data were reprocessed
with CIAO 4.1 using CALDB 4.1.0 and corrected for known time-dependent
gain problems following techniques similar to those described in the
{\it Chandra} analysis
threads\footnote{http://cxc.harvard.edu/ciao/threads/index.html}.  No
charge transfer inefficiency (CTI) correction is available for data
taken at $-110^{\circ}$C.  Screening of the event files was applied to
filter out strong background flares.  Blank-sky background files,
filtered in the same manner as in the HCG 62 image and normalized to
the count rate of the source image in the 10.0-12.0 keV band, were used
for background subtraction.  The final exposure time is 46.5 ks. We
only use data from the S3 CCD since it covers the central part of the
group emission where the cavity system and the radio source are
located.  We identified and removed the point sources on S3 using the
CIAO task {\ttfamily WAVDETECT}, with the detection threshold set to
the default value of $10^{-6}$.

Updated gain correction at $-110^{\circ}$C only become available with
CIAO 4.0 CALDB version, so previous work has been limited by calibration
uncertainties. In particular, spectra were extracted using the
{\ttfamily SPECEXTRACT} task and, as recommended for $-110^{\circ}$C
data\footnote{ http://cxc.harvard.edu/ciao/why/110.html}, new RMF
files were computed using the task {\ttfamily MKACISRMF} with the
latest gain file, {\ttfamily acisD2000-01-29p2\_respN0005.fits}.
\newline

\subsection{{\it XMM--Newton}}

HCG 62 was observed by \textit{XMM--Newton} in June 2007 during
revolution 1382 (ObsID 0504780501, hereafter \#501) and 1383 (ObsID
0504780601, hereafter \#601) for nominal exposure times of 122.5 ks
and 32 ks, respectively.  In this paper, only data from the European
Photon Imaging Camera (EPIC) are presented.  The MOS detectors were
operated in Full Frame Mode and the PN detector in Extended Full Frame
Mode, both with the MEDIUM filter.  We used the SASv8.0.0 processing
tasks {\ttfamily emchain} and {\ttfamily epchain} to generate
calibrated event files from raw data. Throughout this analysis single
and double pixel events for the PN data (PATTERN $<$=4) were selected,
while for the MOS data sets the PATTERNs 0-12 were used. The removal
of bright pixels and hot columns was done by applying the expression
(FLAG==0). To reject soft proton flares, we generated a light
curve in the 10.0-12.0 keV band where the emission is dominated by the
particle--induced background, and excluded all the intervals of
exposure time having a count rate higher than a certain threshold
value (the chosen threshold values are 0.20 (0.18) cps for MOS and
0.35 (0.32) cps for PN in obs. \#501 (\#601)). The total remaining
exposure times after cleaning are 112.0 ks for MOS1, 97.6 ks for MOS2
and 60.2 ks for PN. Starting from the output of the SAS detection
source task, we made a visual selection on a wide energy band MOS \&
PN image of point sources in the field-of-view. Events from these
regions were excluded directly from each event list. The source and
background events were corrected for vignetting using the weighted
method described in Arnaud et al. (2001), the weight coefficients
being tabulated in the event list with the SAS task {\ttfamily
  evigweight}.  This allows us to use the on-axis response matrices
and effective areas, computed in the central 30$''$ with the tasks
{\ttfamily rmfgen} and {\ttfamily arfgen}.

\subsubsection{Background treatment}

The background estimates are derived using blank-sky observations
consisting of a superposition of pointed observations that have been
processed with SAS version 7.1.0 (Carter and Read 2007). In
particular, for each camera, we obtained a blank-sky template tailored
to the characteristics of HCG 62 specific observation by submitting a
XMM-Newton EPIC Background Blank Sky Products Request Form\footnote{
http://www.star.le.ac.uk/~jac48/BG/UserRequest/blankskyform.html}
with the request\footnote{The selection criteria adopted are: Filter:
  Medium, Model: Full-Frame (Extended Full-Frame for PN), Type:
  Ghosted, Galactic column: $2.0-5.0\times 10^{20} {\rm cm}^{-2}$,
  Equatorial coordinates: (193.274,-9.2036) degrees, radius = 180
  degrees.}  of a Galactic column in the range 2 to 5 $\times 10^{20}
{\rm cm}^{-2}$.

The blank-sky background events were then selected using the same
selection criteria (such as PATTERN, FLAG, etc.), intensity filter
(for flare rejection) and point source removal used for the
observation events. This yields final exposure times for the blank
fields of 1.0 Ms for MOS1, 1.6 Ms for MOS2 and 99.3 ks for PN. Since
the cosmic ray induced background might change slightly with time, we
computed the ratio of the total count rates in the high energy band
(10.0-12.0 keV). The obtained normalization factors (1.258 (1.203),
1.316 (1.246), 1.353 (1.175) for MOS1, MOS2 and PN, respectively, in
obs. \#501 (\#601)) were then used to renormalize the blank field
data. The blank-sky background files were finally recast in order to
have the same sky coordinates as HCG 62.

The usual approach to perform the background subtraction is described
in full detail in Arnaud et al. (2002). This procedure consists of two
steps. In the first step, for each spectrum extracted from the
observation event list an equivalent spectrum is extracted from the
corresponding blank-field file and then subtracted from it. This
allows us to remove the particle background. However, if the
background in the observation region is different from the average
background in blank field data, this step could leave a residual
background component. The residual background component is estimated
by using blank-field-subtracted data in a region supposedly free of
source emission (in particular we considered an annulus lying between
13 and 14 arcmin) and then subtracted in a second step from each MOS
and PN spectrum.  We note that while some group emission might still
fall in the annulus chosen to estimate the residual background
component, our spectral analysis is not affected as the central region
we are interested in is very bright and source dominated.

\subsection{{\it GMRT}}

\begin{figure*}[ht]
\centerline{
\includegraphics[width=8.5cm]{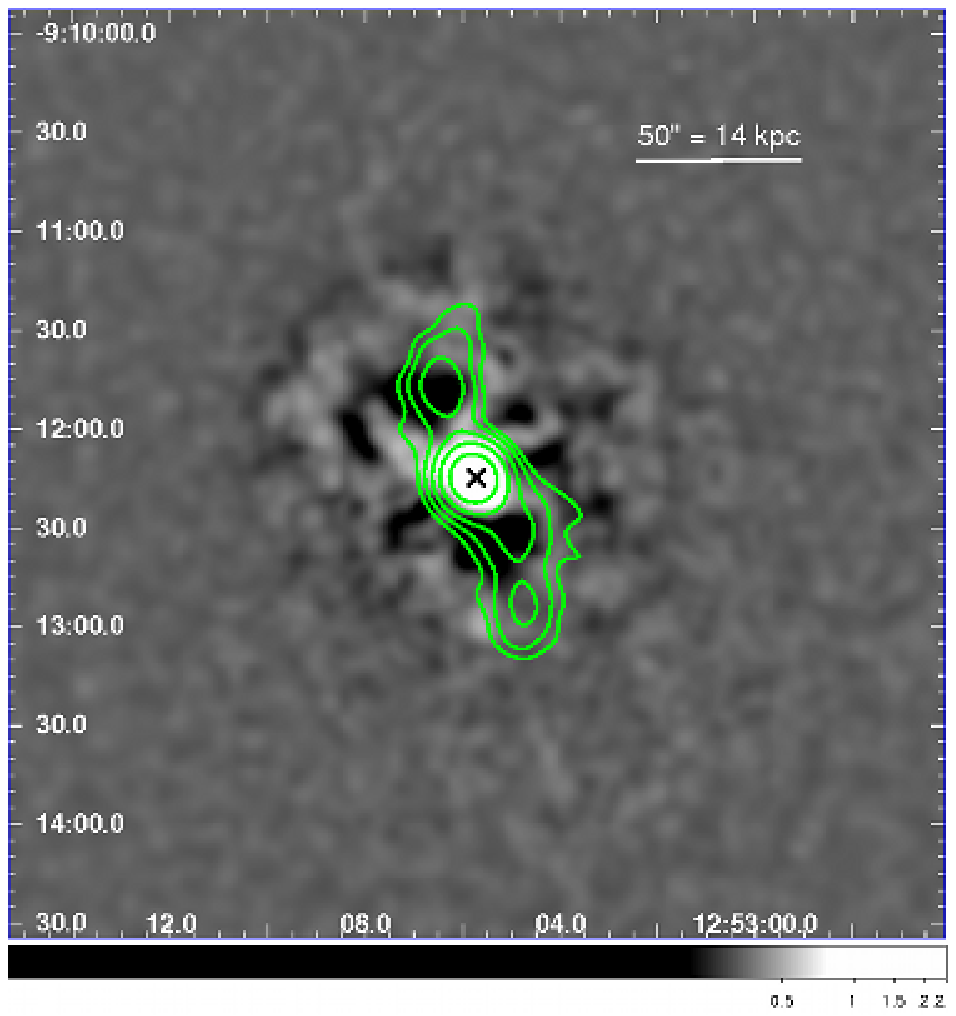}
\includegraphics[width=8.3cm]{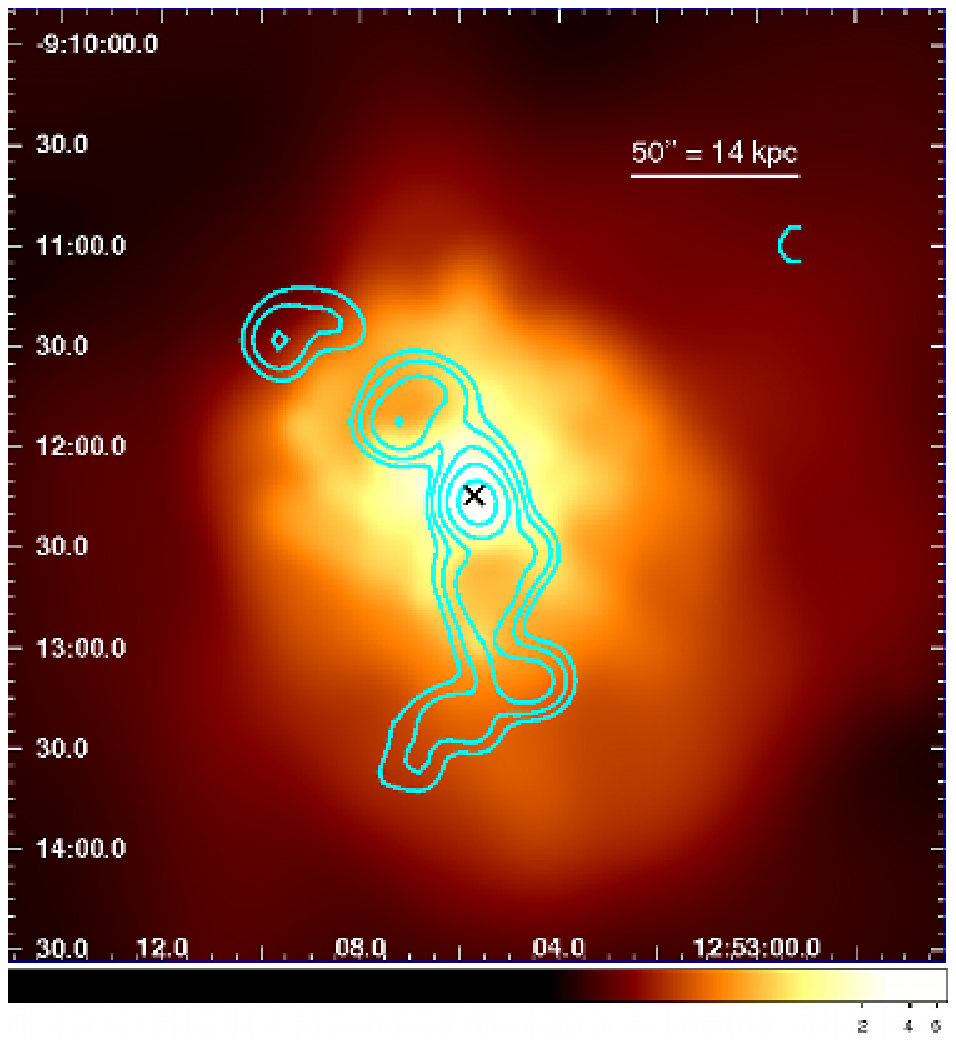}
}
\caption{\label{overlay.fig} \textit{Left:} {\it GMRT} 610 MHz
  contours overlaid on the unsharp masked 0.5-2.0 keV {\it Chandra}
  image (same as in top right of Fig. \ref{xmm-chandra.fig}).  The
  beam size is $14''$ by $14''$ and the lowest contour is shown at $3
  \sigma = 0.15$ mJy/beam.  \textit{Right:} {\it GMRT} 235 MHz
  contours overlaid on the smoothed 0.5-2.0 keV {\it Chandra} image.
  The beam size is $14.4''$ by $13''$ and the r.m.s. noise is $170
  \mu$Jy/beam. The contours are spaced by a factor 2 starting from the
  lowest level of 0.45 mJy/beam. In both panels, the cross indicates
  the position of the radio core. North is up and east is left. }
\end{figure*}

HCG 62 was observed by the {\it GMRT} in February 2008 at frequencies
of 610 MHz and 235 MHz (project 13SGa01) for a total of 125 and 250
minutes, respectively. The data were acquired in spectral line mode
with 128 channels/band and a bandwidth of 16 MHz at 610 MHz and 8 MHz
at 235 MHz.  Data reduction was done using the NRAO AIPS (Astronomical
Image Processing System) package. Accurate editing of the uv data was
applied to identify and remove bad data. After bandpass calibration,
the central channels were averaged to 6 channels of $\sim$ 2 MHz at
610 MHz and $\sim$ 1 MHz at 235 MHz. After further careful editing in
the averaged datasets, images were produced using the standard
procedure (calibration, Fourier-Transform, Clean and Restore) and the
wide-field imaging technique. Phase-only self-calibration was applied
to remove residual phase variations. The rms noise level ($1\sigma$)
in the final images is 50 $\mu$Jy/beam at 610 MHz and 170 $\mu$Jy/beam
at 235 MHz.  Following Chandra et al. (2004), we estimated that the
amplitude calibration error at 610 MHz is $\sim 5\%$, while a
calibration error of 8\% was assumed at 235 MHz.
The calibration uncertainties were included in the flux density
and spectral index errors determined in \S \ref{radio.sec}.
The {\it GMRT} observations of HCG 62 will be presented and discussed
in more detail in Giacintucci et al. (in prep.).


\section{The disturbed inner region of HCG 62}
\label{disturbed.sec}

\subsection{X-ray morphology}
\label{X-morphology.sec}

The raw 0.5-2.0 keV ACIS-S image in Fig. \ref{xmm-chandra.fig} (top
left panel) shows that the hot gas in the central region of HCG 62 is
not smoothly distributed but is instead perturbed with cavities and
edges. In particular, as best revealed by the unsharp masked image
shown in Fig. \ref{xmm-chandra.fig} (top right panel), we confirm the
presence of two clear cavities, first discovered by Vrtilek et
al. (2002), and we also notice the presence of surface brightness
discontinuities or ``fronts'' which suggest the presence of
large-scale gas motions or shocks in the group halo.  In particular,
by means of the surface brightness profiles presented in \S
\ref{front.sec}, we identify a front at 36 kpc ($\sim 2'$) from the
center toward the south-west (SW) at a distance of about 10 kpc ($\sim
35''$) from the southern radio lobe, and one at 20 kpc ($\sim 70''$)
from the center toward the north-east (NE) close to the N cavity.
These features are significant at the 5$\sigma$ and 15$\sigma$ level,
respectively.  In the images there is also a hint of an edge about the
same distance SW of the group center as the front to the NE.  The
spectral analysis presented below indicates that the outer SW edge is
a shock front (\S \ref{front.sec}).  As already estimated by previous
studies (B\^irzan et al. 2004, Rafferty et al. 2006, Morita et
al. 2006) the two cavities have similar sizes, with diameters of $\sim
10$ kpc, and are located at a projected distance of $\sim 8.5$ kpc
from the group center. We extracted the surface brightness profiles
along and orthogonal to the cavities, and at the radius of $\sim 8.5$
kpc we estimate a brightness decrement of $\sim$ 30\% along the
cavities relative to the orthogonal direction.

Because of the poorer {\it XMM--Newton} spatial resolution, the
cavities are not evident in the raw 0.5-2.0 keV image shown in
Fig. \ref{xmm-chandra.fig} (bottom left panel). However, they do
appear clearly in the unsharp masked image
(Fig. \ref{xmm-chandra.fig}, bottom right panel), at positions
coincident with those determined by the {\it Chandra} data.

\subsection{Radio properties}
\label{radio.sec}

The existing higher frequency (1.4 GHz) observations of HCG 62 show
the central compact radio source with some hint of emission extending
toward the southern cavity (Vrtilek et al. 2002).  With the new {\it
  GMRT} observations at 235 MHz and 610 MHz we detect clearly extended
emission emanating from the core in the typical form of a bipolar
flow. The two radio lobes point toward the N and S directions, filling
the cavities.  The overlay of the 610 MHz radio contours on the
unsharp masked {\it Chandra} image is shown in Fig. \ref{overlay.fig}
(left), whereas the overlay of the 235 MHz radio contours on the
smoothed {\it Chandra} image is shown in Fig. \ref{overlay.fig}
(right). It is evident that far more extensive structures become
visible at lower frequencies: the radio emission at 235 MHz is more
extended with two faint regions located outside the cavities and
apparently bent toward the east (E).  In particular, we identify inner
lobes clearly filling the well defined X-ray cavities (best visible
in left panel of Fig.  \ref{overlay.fig}), and outer lobes having no
associated X-ray cavities (best visible in right panel of Fig.
\ref{overlay.fig}).  The connection of the radio source with the X-ray
features will be discussed in \S \ref{discussion.sec}.

At 610 MHz, the source has a total flux density of $13.5 \pm 0.7$ mJy.
At 235 MHz, the total flux density is $42.4 \pm 3.4$ mJy.  Neglecting
the {\it K-correction} term (for 
$0.3 < \alpha < 1.7$ the {\it K-correction} is $<1$\%, 
Petrosian \& Dickey 1973), the
monochromatic radio power at each frequency is calculated as
\begin{equation}
  P_{\nu}=4 \pi \, D_{\rm L}^2 \, S_{\nu}
\label{radio_power.eq}
\end{equation}
where $D_{\rm L}$ is the luminosity distance, which gives $P_{\rm 610
  \, MHz} = (5.7 \pm 0.3) \times 10^{21}$ W Hz${^{-1}}$ and $P_{\rm
  235 \, MHz} = (1.8 \pm 0.1) \times 10^{22}$ W Hz${^{-1}}$.  This
radio source is thus classified as a weak FR-I, and its spectral index
$\alpha^{\rm 610 \, MHz}_{\rm 235 \, MHz} = 1.2 \pm 0.1$ is steep
compared to that of typical radio galaxies.

Since the outer radio lobes are only detected at one frequency in our
{\it GMRT} observations, we are unable to perform a detailed radio
spectral analysis.  New, deeper {\it GMRT} observations at 235, 327
and 610 MHz, scheduled during Cycle 17, will allow us to study with
unprecedented detail the spectral properties of the source. Based on
the analysis of the integrated radio spectrum and spectral index
distribution, an estimate of the radiative ages of the electron
population can be obtained (e.g., Giacintucci et al.  2007, 2008 and
references therein) for both pairs of radio lobes. This will help to
shed light on the relationship of the outer lobes to the inner ones.


\begin{figure*}[ht]
\centerline{
\includegraphics[width=8.5cm]{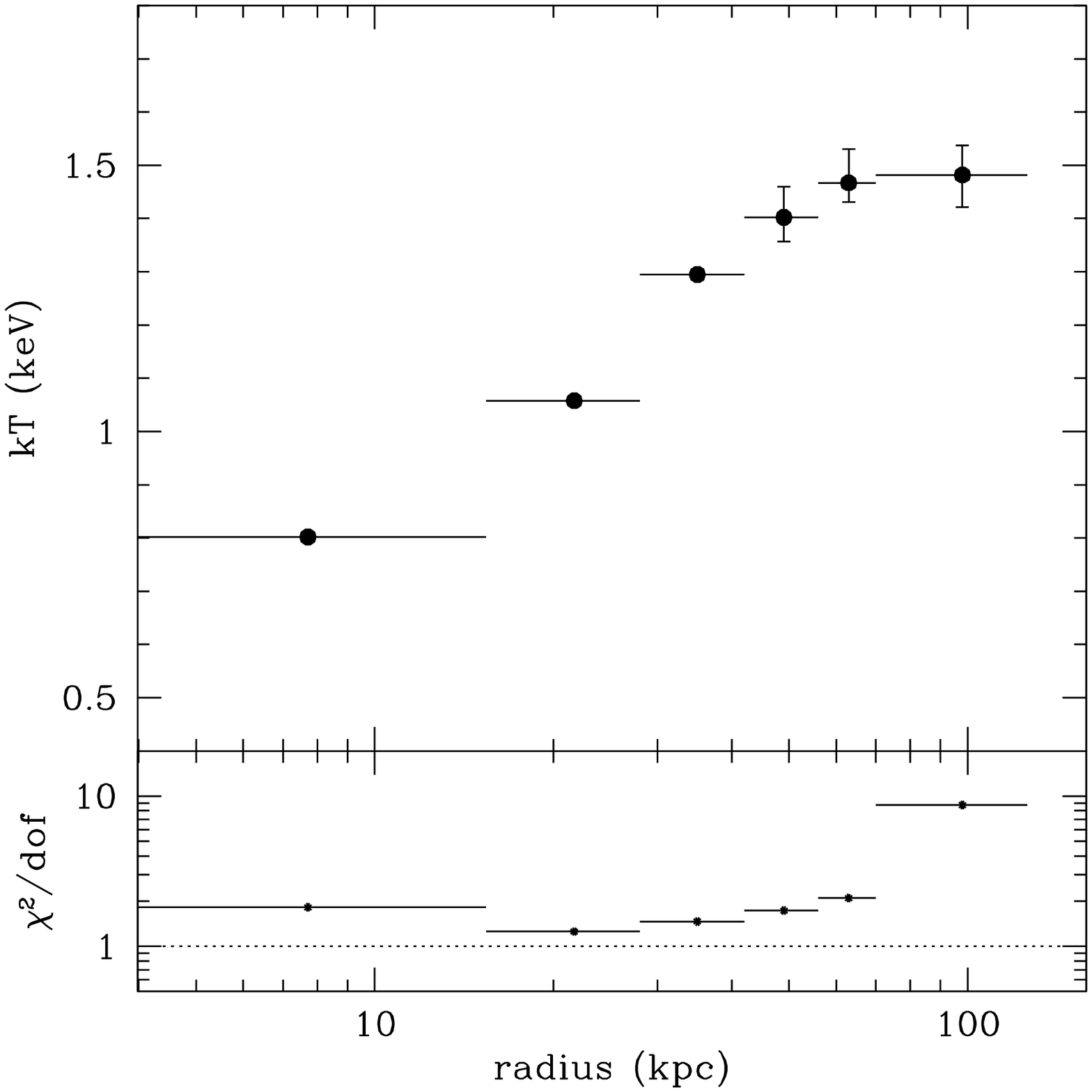}
\includegraphics[width=8.5cm]{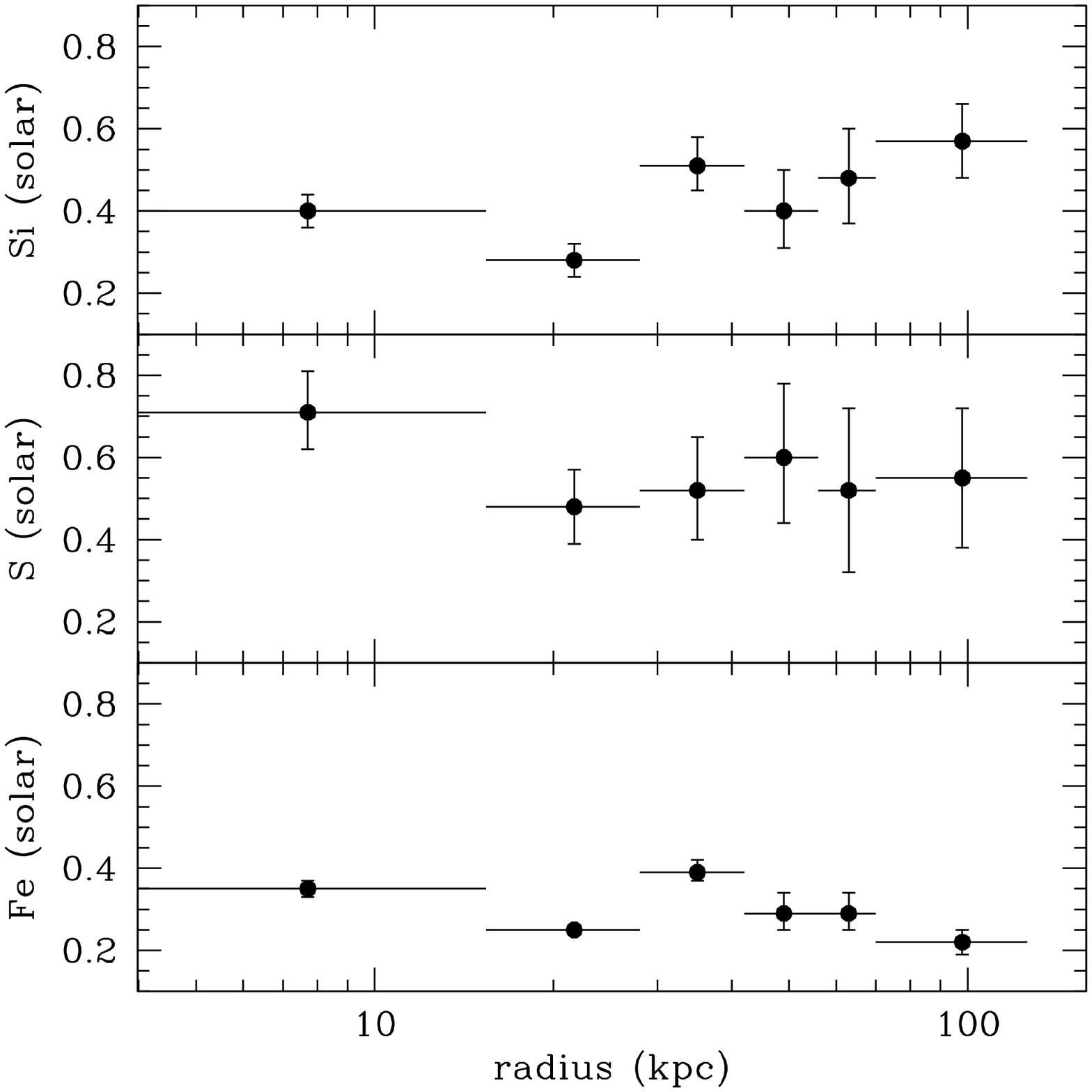}
}
\caption{\label{profiles-proj.fig} \textit{Left:} Azimuthally averaged
  projected temperature profile measured with {\it XMM--Newton} and
  reduced $\chi^2_{\nu} = \chi^2/{\rm dof}$ of the fit.
  \textit{Right:} Azimuthally averaged projected abundance profiles of
  Si, S, and Fe measured with {\it XMM--Newton}.  }
\end{figure*}


\begin{deluxetable}{ccccccc}[ht]
\tablewidth{0pt}
\tablecaption{{\it XMM--Newton} Projected Spectral Analysis
\label{profile_proj.tab}
}
\tablehead{
\colhead{\#} & \colhead{$kT$} &  \colhead{Mg,Ne,O}  &  \colhead{Si}  &  \colhead{S}  &  \colhead{Ni,Fe} & \colhead{$\chi^2$/dof}\\
\colhead{} & \colhead{(keV)} & \colhead{(solar)}  & \colhead{(solar)}  & \colhead{(solar)}  & \colhead{(solar)} & }
\startdata
1 & $0.80^{+0.01}_{-0.01}$ & $0.42^{+0.06}_{-0.05}$
 & $0.40^{+0.04}_{-0.04}$  & $0.71^{+0.10}_{-0.09}$  & $0.35^{+0.02}_{-0.02}$
& 1430/786
\\[+1mm]
2 & $1.06^{+0.01}_{-0.01}$ & $0.08^{+0.07}_{-0.07}$
 & $0.28^{+0.04}_{-0.04}$  & $0.48^{+0.09}_{-0.09}$  & $0.25^{+0.01}_{-0.01}$
& 820/653
\\[+1mm]
3 & $1.30^{+0.01}_{-0.01}$ & $0.00^{+0.06}_{-0.00}$
 & $0.51^{+0.07}_{-0.06}$  & $0.52^{+0.13}_{-0.12}$  & $0.39^{+0.03}_{-0.02}$
& 935/638
\\[+1mm]
4 & $1.40^{+0.06}_{-0.05}$ & $0.14^{+0.20}_{-0.06}$
 & $0.40^{+0.10}_{-0.09}$  & $0.60^{+0.18}_{-0.16}$  & $0.29^{+0.05}_{-0.04}$
& 1000/578
\\[+1mm]
5 & $1.47^{+0.06}_{-0.04}$ & $0.00^{+0.14}_{-0.00}$
 & $0.48^{+0.12}_{-0.11}$  & $0.52^{+0.20}_{-0.20}$  & $0.29^{+0.05}_{-0.04}$
& 1258/600
\\[+1mm]
6 & $1.48^{+0.06}_{-0.06}$ & $0.00^{+0.02}_{-0.00}$
 & $0.57^{+0.09}_{-0.09}$  & $0.55^{+0.17}_{-0.17}$  & $0.22^{+0.03}_{-0.03}$
& 10273/1178
\\[-2mm]
\enddata
\tablecomments{ Results of the {\it XMM--Newton} spectral fitting in
  concentric $360^{\circ}$ annular regions in the 0.7-3.0 keV energy
  range using the absorbed {\ttfamily vapec} thermal plasma model and
  fixing the absorbing column density to the Galactic value ($N_{\rm
    H} = 3.31 \times 10^{20} {\rm cm}^{-2} $).  The temperature (in
  keV) and abundances (in fraction of the solar value, Anders \&
  Grevesse 1989) are left as free parameters. Error bars are at the
  90\% confidence levels on a single parameter of interest.  The first
  column indicate the number region. The delimiting radii are:
  reg.\#1: 0-15 kpc (0-55$''$), reg.\#2: 15-28 kpc (55$''$-100$''$),
  reg.\#3: 28-42 kpc (100$''$-150$''$), reg.\#4: 42-56 kpc
  (150$''$-200$''$), reg.\#5: 56-70 kpc (200$''$-250$''$), reg.\#6:
  70-126 kpc (250$''$-450$''$).  }
\end{deluxetable}


\section{Global X-ray properties}
\label{xray.sec}

Throughout the analysis, a single spectrum for each instrument and
each dataset was extracted for each region of interest.  The relative
normalizations of the MOS and PN spectra were left free when fitted
simultaneously. Spectral fitting was performed in XSPEC version
12.3.1. Abundances were measured relative to the abundance ratios of
Anders \& Grevesse (1989).  A galactic hydrogen column of $3.31 \times
10^{20} {\rm cm}^{-2}$ (Dickey \& Lockman 1990) and a redshift of
0.0137 was assumed and, unless otherwise stated, the reported errors
are at the 90\% confidence level.  Spectra were grouped to 25 counts per
bin and the energy range 0.7-3.0 keV was adopted.

We first tested in detail the consistency between {\it Chandra} and
the three {\it XMM--Newton} cameras by extracting a global spectrum
from all events lying in the annulus 1-3 arcmin from the group
emission peak and fitting these spectra separately with an absorbed
{\ttfamily apec} model.  We find good agreement between the three {\it
  XMM--Newton} cameras ($kT = 1.25^{+0.02}_{-0.02}$ keV for MOS1,
$1.21^{+0.02}_{-0.02}$ keV for MOS2, $1.23^{+0.01}_{-0.01}$ keV for
PN, with a value of $1.23^{+0.01}_{-0.01}$ for the combined fit),
whereas there is a discrepancy of about 0.1 keV with the {\it Chandra}
measurement ($kT = 1.32^{+0.01}_{-0.01}$ keV).  However, this is
consistent with the systematic cross-calibration uncertainties (David
2009).  Due to the systematic offset, we analyzed {\it Chandra} and
{\it XMM--Newton} data separately. In particular, we present here the
results obtained by the analysis of {\it XMM--Newton} spectra only
because they comprise most of the counts.

\subsection{Azimuthally Averaged Projected Temperature and Metallicity 
Profiles}
\label{averaged.sec}

We produced projected radial temperature and metallicity profiles by
extracting spectra in six circular annuli centered on the peak of the
X--ray emission. The annular regions are described in Table
\ref{profile_proj.tab}.  The data from the three {\it XMM--Newton}
cameras were fitted simultaneously to an absorbed {\ttfamily vapec}
thermal plasma model, where O, Si, S, and Fe are treated as free
parameters with the Ni abundance linked to the Fe abundance and the Mg
and Ne abundances linked to the O abundance.  The best-fitting
parameter values and 90\% confidence levels derived from the fits to
the $360^{\circ}$ annular spectra are summarized in Table
\ref{profile_proj.tab}.

The azimuthally averaged projected temperature profile
(Fig. \ref{profiles-proj.fig} left) shows a positive temperature
gradient at small radii, with temperatures increasing up to $\sim 1.5$
keV at a radius of $\sim$ 50 kpc.  This behavior is similar to that
observed in other groups (e.g., Gastaldello et al. 2007; Finoguenov et
al. 2007; Rasmussen \& Ponman 2007; Sun et al. 2009; O'Sullivan et
al. in prep.), although we do not observe the typical decline in
temperature at larger radii. Previous {\it ROSAT} observations of HCG
62 indicate a temperature decline beyond $\sim$ 120 kpc (Ponman \&
Bertram 1993), so our result is consistent with earlier literature.

The azimuthally averaged projected abundance profiles (Fig.
\ref{profiles-proj.fig}, right) do not show strong gradients, but are
instead relatively flat around values of $\sim$0.5 solar for both Si
and S, and $\sim$0.3 solar for Fe. We were able to constrain O only
in the central region, with a value of $\sim$0.4 solar.


\begin{deluxetable}{cccccc}
\tablewidth{0pt}
\tablecaption{{\it XMM--Newton} Deprojected Spectral Analysis
\label{profile_deproj.tab}
 }
\tablehead{
\colhead{\#} & \colhead{$kT$} &  \colhead{Ni,Fe}  &  \colhead{$n_{\rm e} \times 10^{-3}$} & \colhead{$P \times 10^{-12}$} & \colhead{$S$} \\
\colhead{} & \colhead{(keV)} & \colhead{(solar)}  & \colhead{$({\rm cm}^{-3})$} & \colhead{$({\rm erg \, cm}^{-3})$}  & \colhead{$({\rm keV \, cm}^2)$} 
}
\startdata
1 & $0.36^{+0.01}_{-0.01}$ & --
 & $26.78^{+3.54}_{-3.99}$ & $30.89^{+4.09}_{-4.61}$  & $4.02^{+0.36}_{-0.40}$
\\[+1mm]
2 & $1.00^{+0.01}_{-0.01}$ & $0.41^{+0.03}_{-0.02}$
 & $3.57^{+1.09}_{-0.62}$ & $11.43^{+3.48}_{-2.00}$  & $42.83^{+8.71}_{-4.99}$
\\[+1mm]
3 & $1.28^{+0.02}_{-0.02}$ & $0.52^{+0.06}_{-0.05}$
 & $2.00^{+0.36}_{-0.43}$ & $8.19^{+1.47}_{-1.76}$  & $80.72^{+9.69}_{-11.59}$
\\[+1mm]
4 & $1.33^{+0.11}_{-0.04}$ & $0.32^{+0.07}_{-0.04}$
 & $1.14^{+0.38}_{-0.43}$ & $4.88^{+1.66}_{-1.82}$  & $121.53^{+28.52}_{-30.31}$
\\[+1mm]
5 & $1.55^{+0.10}_{-0.12}$ & $0.45^{+0.13}_{-0.12}$
 & $0.89^{+0.22}_{-0.30}$ & $4.43^{+1.14}_{-1.50}$  & $167.32^{+29.75}_{-39.03}$
\\[+1mm]
6 & $1.47^{+0.06}_{-0.06}$ & $0.21^{+0.04}_{-0.03}$
 & $0.59^{+0.09}_{-0.12}$ & $2.79^{+0.43}_{-0.59}$  & $208.34^{+22.15}_{-29.87}$
\\[-2mm]
\enddata
\tablecomments{Results of the {\it XMM--Newton} deprojection analysis in
  concentric $360^{\circ}$ annular regions in the 0.7-3.0 keV energy
  range using the XSPEC {\ttfamily projct$\times$wabs$\times$mekal}
  model and fixing the absorbing column density to the Galactic value
  ($N_{\rm H} = 3.31 \times 10^{20} {\rm cm}^{-2} $). The fit gives
  $\chi^2$/dof = 19848/4463.  The first column indicate the number
  region. The delimiting radii are the same as in Table
  \ref{profile_proj.tab}.   }
\end{deluxetable}


\subsection{Deprojection Analysis}
\label{deproj.sec}

\begin{figure*}[ht]
\centerline{
\includegraphics[width=8.5cm]{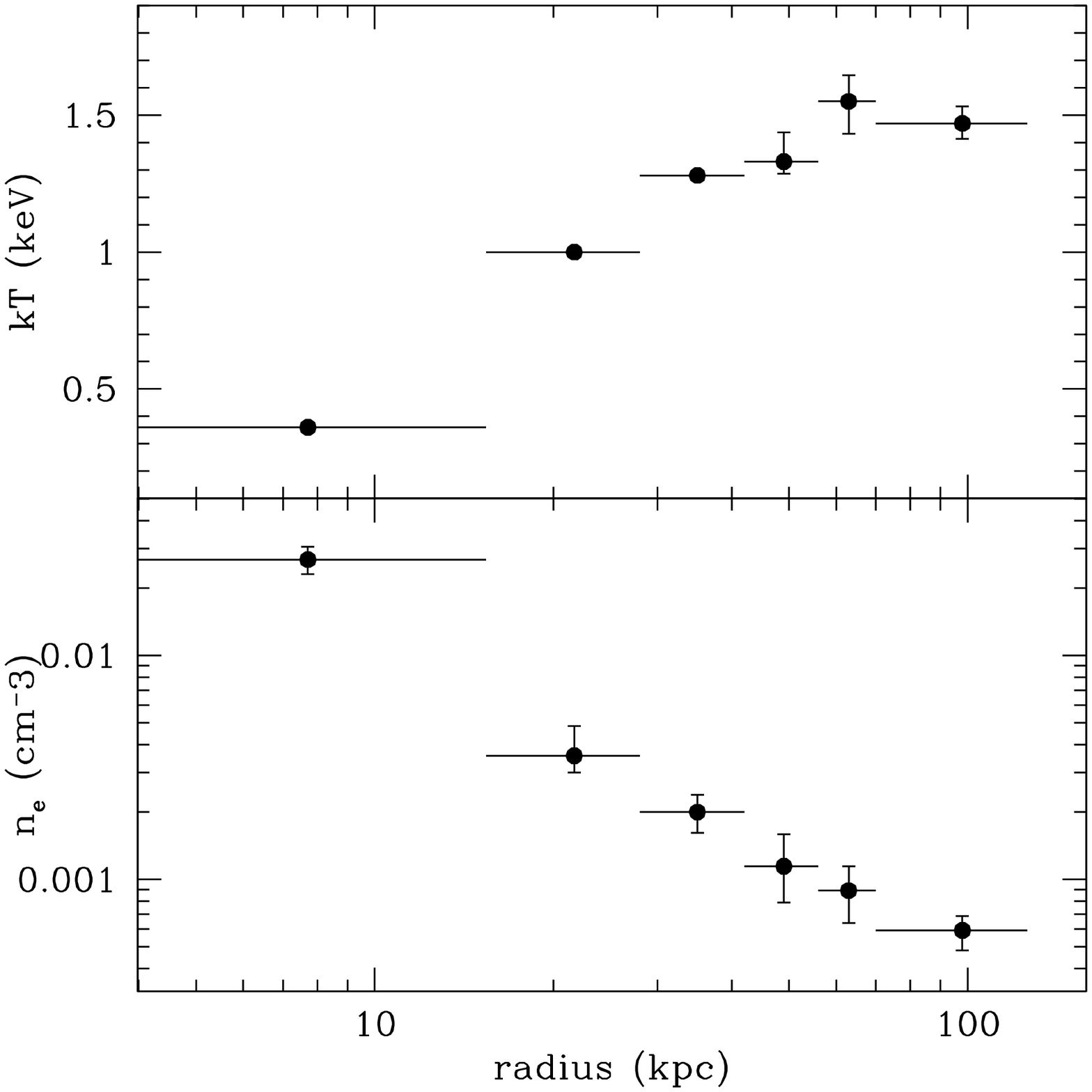}
\includegraphics[width=8.5cm]{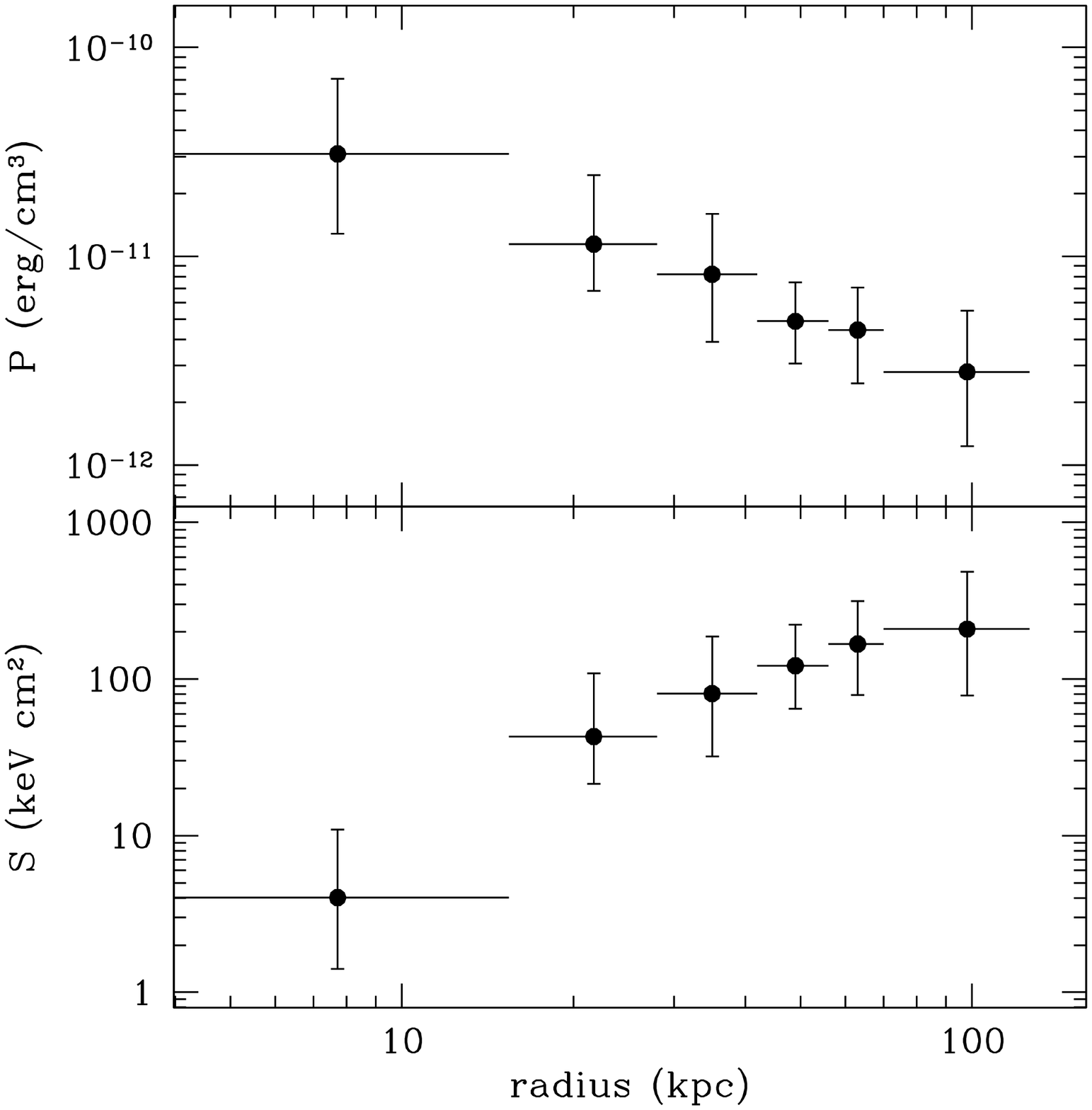}
}
\caption{\label{profiles-deproj.fig}
\textit{Left:} Azimuthally averaged deprojected profiles of temperature 
(top) and density (bottom) measured with {\it XMM--Newton}. 
\textit{Right:}
Azimuthally averaged deprojected profiles of pressure (top) and 
entropy (bottom) measured with {\it XMM--Newton}. }
\end{figure*}

\begin{figure*}[ht]
\centerline{
\includegraphics[width=8.5cm]{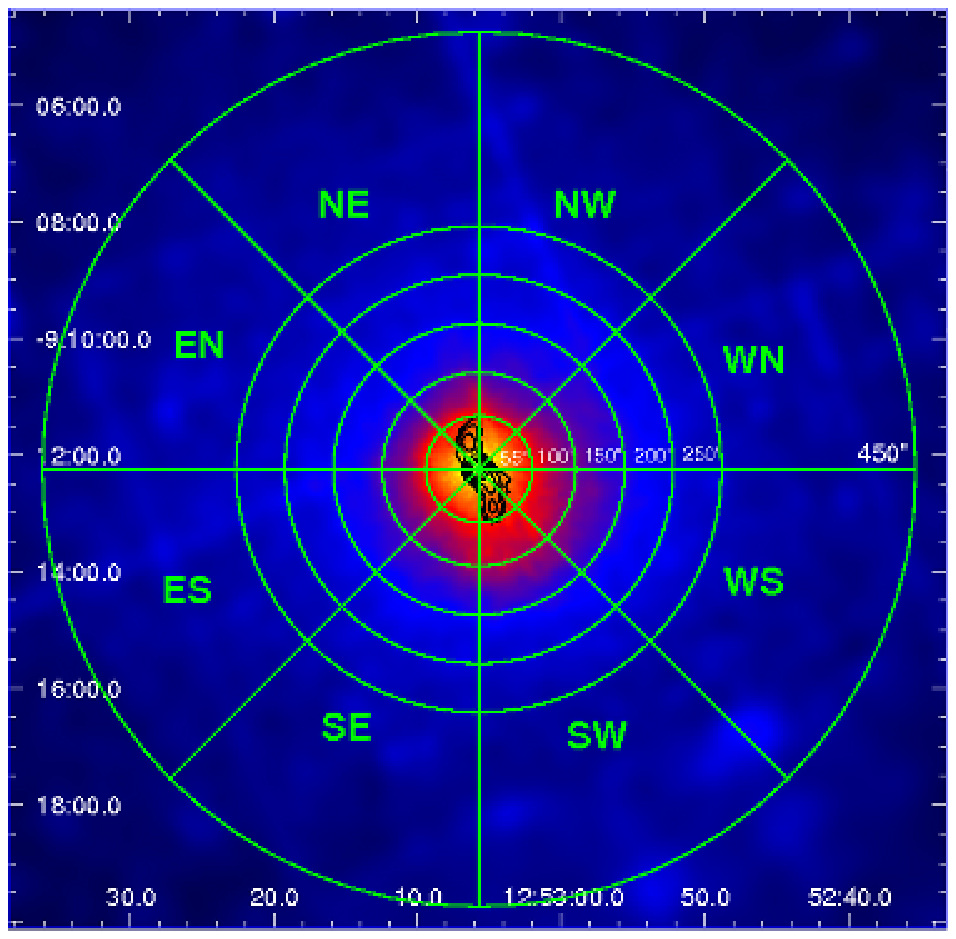}
\includegraphics[width=8.5cm]{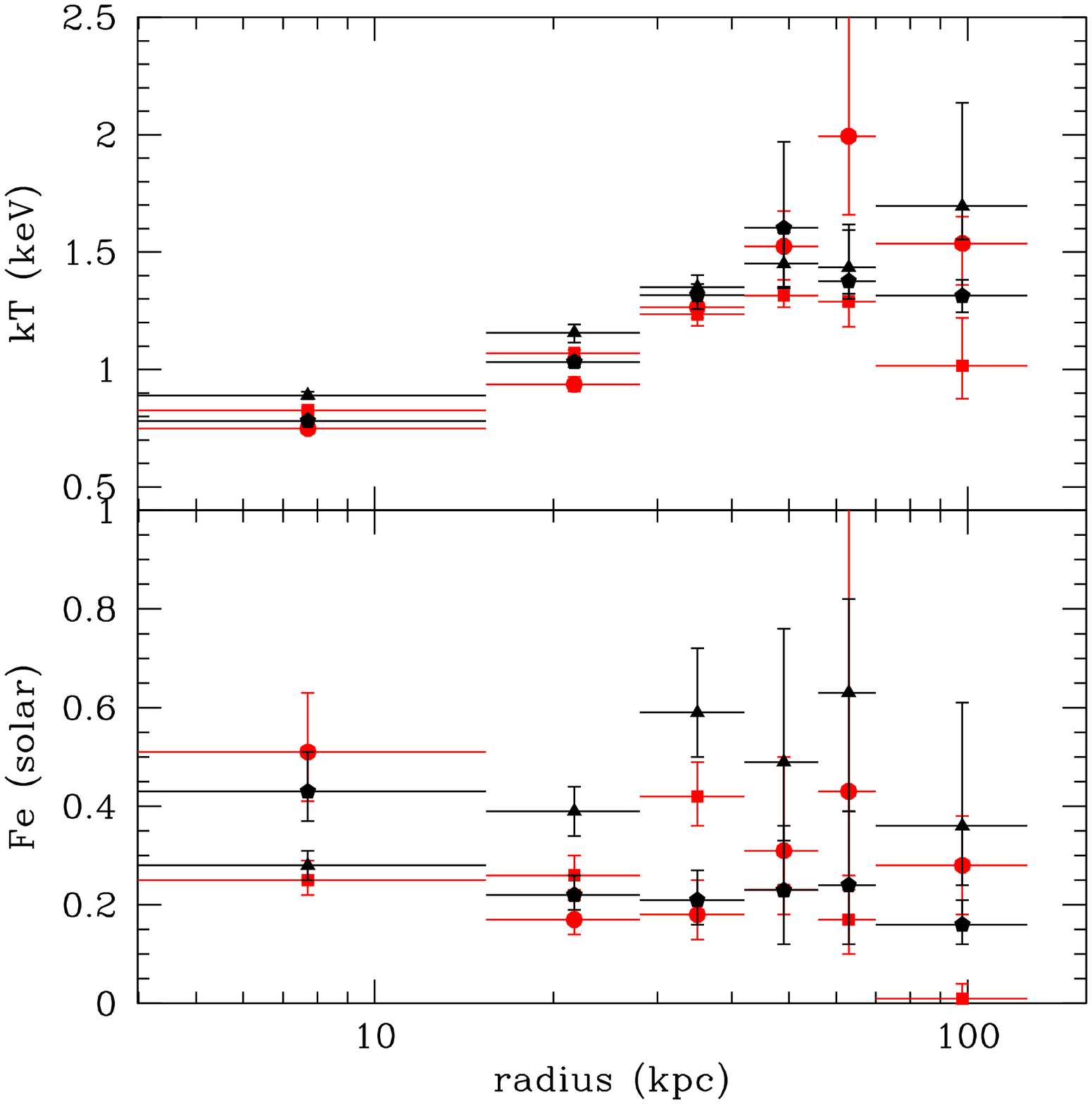}
}
\centerline{
\includegraphics[width=8.5cm]{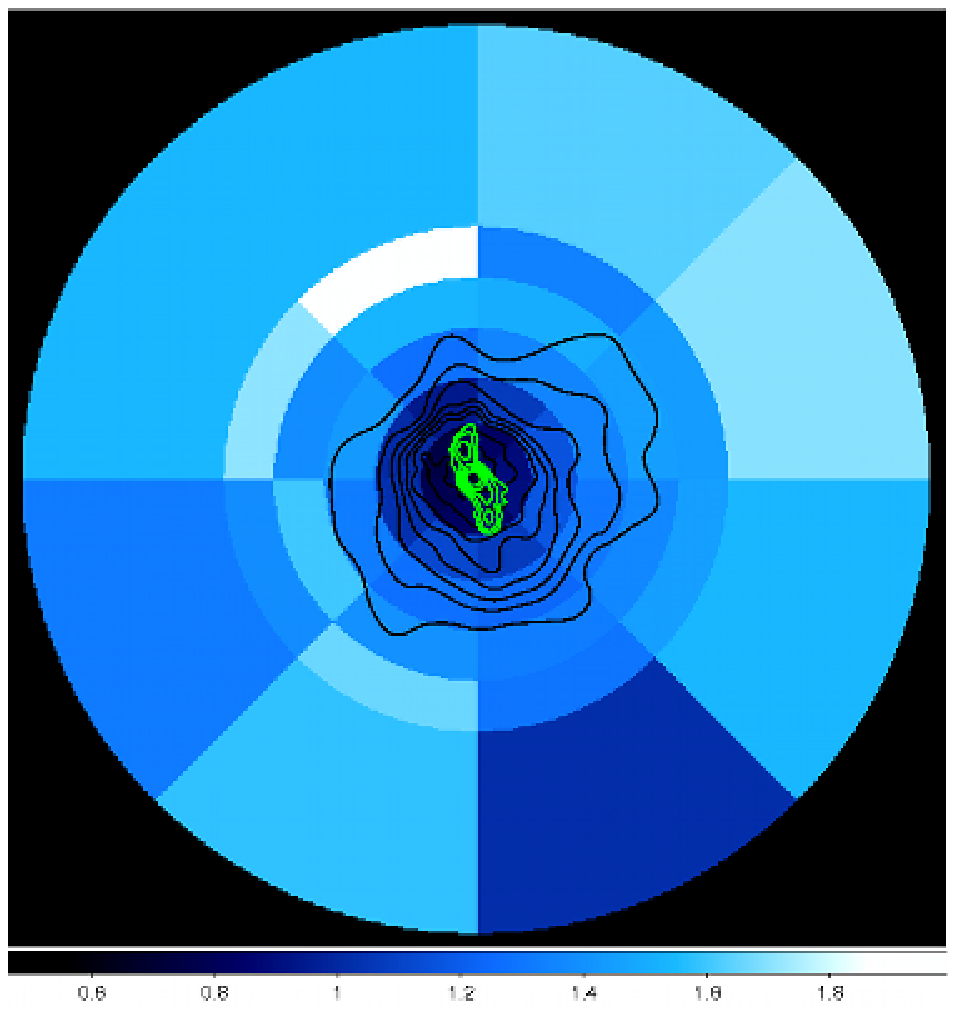}
\includegraphics[width=8.5cm]{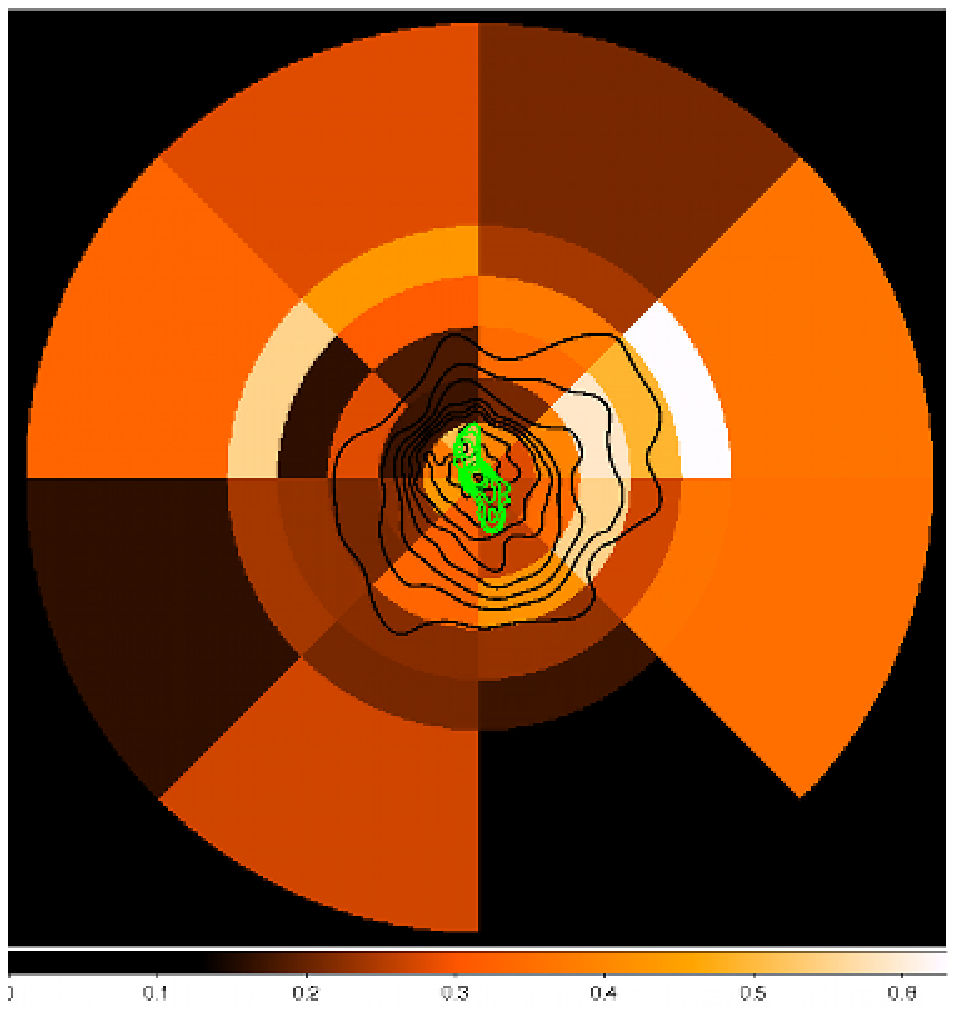}
}
\caption{\label{sectors-proj.fig} \textit{Top, left: } Overlaid on the
  smoothed 0.5-2.0 keV {\it XMM--Newton} mosaic image are the sectors
  used for the spectral analysis (see \S \ref{azimuthal.sec} for
  details).  610 MHz {\it GMRT} radio contours are also shown to
  locate the position of the inner radio lobes.  \textit{Top, right:}
  Projected temperature (top panel) and Fe abundance (bottom panel)
  profiles measured with {\it XMM--Newton} along sector NE (red
  circles), sector SW (red squares), sector WN (black triangles) and
  sector ES (black pentagons).  In red and black are indicated the
  profiles measured along and orthogonal to the cavities,
  respectively.  \textit{Bottom:} ``Wedge'' temperature ({\it left})
  and Fe abundance ({\it right}) maps derived from the {\it
    XMM--Newton} projected spectral fits in the sectoral annuli
  detailed in top, left figure. For comparison, {\it Chandra} contours
  and 610 MHz {\it GMRT} contours are overlaid in black and green,
  respectively.}
\end{figure*}

To correct for the effect of projection along the line of sight, we
also performed a deprojection analysis on the same annular spectra
used in \S \ref{averaged.sec} by adopting the XSPEC {\ttfamily projct}
model. Under the assumption of ellipsoidal (in our specific case,
spherical) shells of emission, this model calculates the geometric
weighting factor, according to which the emission is redistributed
amongst the projected annuli.  As expected, the deprojected central
temperature is lower than the projected one, since in the projected
fits the spectrum of the central annulus is contaminated by hotter
emission along the line of sight. We also note that the low value that
we measure probably represents the very coolest gas in the core, owing
to emission weighting in the large bin adopted to account for the {\it
  XMM} PSF.  Due to the limited photon statistics relative to the high
number of parameters in the {\ttfamily projct} model we could only
poorly constrain the deprojected abundance profiles, which therefore
are not shown.
 
We also estimate various quantities derived from the deprojected
spectral fits. The electron density $n_{\rm e}$ is obtained from the
Emission Integral $EI = \int n_{\rm e} n_{\rm p} dV$ given by the
{\ttfamily vapec} normalization: $10^{-14} EI / ( 4 \pi [D_{\rm A}
(1+z)]^2 )$. We assume $n_{\rm p} = 0.82 n_{\rm e}$ in the ionized
intra-cluster plasma. By starting from the deprojected density and
temperature values, we can then calculate the gas pressure as $p=nkT$,
where we assume $n = 2 n_{\rm e}$, and the gas entropy from the
commonly adopted definition $ S=kT \, n_{\rm e}^{-2/3}$.  The results
are reported in Table \ref{profile_deproj.tab} and the corresponding
deprojected temperature, density, pressure and entropy profiles are
shown in Fig. \ref{profiles-deproj.fig}.


\subsection{Azimuthal Variations in Projected Temperature and 
Metallicity Profiles}
\label{azimuthal.sec}

The interaction of the radio source with the thermal gas can produce
significant effects on the X-ray properties of the group halo, that
are thus expected to show azimuthal variations.  To investigate this,
we performed a spectral analysis in annular sectors each having
angular aperture of $45^{\circ}$.  Starting from the W direction with
position angle (P.A.) = $0^{\circ}$ and counting counterclockwise, the
sectors are labeled as: WN (west-north), NW (north-west), NE
(north-east), EN (east-north), ES (east-south), SE (south-east), SW
(south-west), WS (west-south).  Each sector was divided in the same
six annular regions used in \S \ref{averaged.sec}, obtaining the
regions shown in Fig. \ref{sectors-proj.fig} (left panel).  We
produced projected radial temperature and metallicity profiles by
extracting spectra in these regions.  In particular, we compared the
radial properties along and orthogonal to the radio lobes and cavity
system.  The northern and southern cavities fall entirely within the
first annulus in the NE and SW sectors, respectively.  The profiles
derived along the different sectors are shown in Fig.
\ref{sectors-proj.fig} (top right), while the temperature and
metallicity wedge maps are shown in Fig. \ref{sectors-proj.fig}
(bottom left and right, respectively).  The gas properties do not
appear to show variations obviously associated with the cavities and
radio lobes. This indicates that the spectral analysis in large
sectors is probably not the most sensitive way to study the 2-D
distribution of temperature and metallicity in such a disturbed system
as HCG 62. Higher-resolution spectral maps are presented and discussed
in \S \ref{maps.sec}.


\section{Discussion}
\label{discussion.sec}

\subsection{Interaction of the Radio Plasma with the Hot Gas}
\label{interaction.sec}

\subsubsection{Energetics}
\label{energetics.sec}

We wish to compare the cavity power -- which is a measure of the
energy injected into the hot gas by the AGN outburst -- with the gas
luminosity inside the cooling radius -- which represents the energy
lost by the hot gas due to X-ray radiation. The total power of the two
cavities can be estimated directly from the X-ray measurement of the
$pV$ work done by the jets in inflating the cavities, and from
measurement of the cavity age (e.g., McNamara \& Nulsen 2007). By
measuring the cavity enthalpy as $ H = 4 pV$ (for a relativistic
plasma) and the cavity age as the sound crossing time $t_s = R/c_s = R
/ \sqrt{\gamma kT / \mu m_{\rm H}}$, where $R$ is the projected
distance from the center of the cavity to the group center, $\gamma =
5/3$, and $\mu = 0.62$, we estimate that the total cavity power is
$P_{\rm cav} = 3.8 \times 10^{42} {\rm \, erg \, s}^{-1}$, which is in
agreement with the results of B\^irzan et al. (2004) and Rafferty et
al. (2006).  In particular, in deriving the sound speed $c_s$, we
adopt the temperature values estimated in the first NE and SW sectoral
annuli, respectively (see Sect. \ref{azimuthal.sec}).  The results are
reported in Table \ref{cavity.tab}.  Given the uncertainties in
estimating the cavity age, for comparison we also adopt the methods
explained in B\^irzan et al.  (2004) which consider the buoyancy-time
$t_{\rm buoy} \sim R / \sqrt{2 g V / S C}$, where $g = G M_{< R} /
R^2$ is the gravitational acceleration, $V$ is the volume of the
cavity, $S$ is the cross-section of the cavity, and $C=0.75$ is the
drag coefficient (Churazov et al. 2001), and the refill-time $t_{\rm
  ref} \sim 2 \sqrt{r / g}$, where $r$ is the radius of the cavity.
We calculate the gravitational acceleration at the distance of the
cavities from the total mass profile derived by Morita et al. (2006),
who estimate $M_{<R \sim 8.5 {\rm kpc}} \sim 2 \times 10^{11} {\rm
  M}_{\odot}$.  We find that the age estimates agree to within a
factor of 2, with the buoyancy times of $\sim 2.0 \times 10^7$ yr
lying in between the sound crossing time of $\sim 1.8 \times 10^7$ yr
and the refill times of $\sim 3.7 \times 10^7$ yr.

We estimate the gas luminosity inside the cooling radius ($r_{\rm
  cool} = 33 \, {\rm kpc}$, Rafferty et al. 2006) as the bolometric
X-ray luminosity, $L_X$, derived from a deprojection analysis of {\it
  XMM--Newton} spectra similar to that described in \S
\ref{deproj.sec}.  We measure $L_X = 1.5 \times 10^{42} {\rm \, erg \,
  s}^{-1}$, which is consistent with the value of Rafferty et
al. (2006).  This indicates that the mechanical luminosity of the AGN
outburst is large enough to balance the radiative losses.  In fact,
since the 235 MHz radio emission extends beyond the cavities, we may
argue that there is more power in the jet and lobes than one can infer
from the X-rays.  The cavity volumes may be larger as the radio
emission is less sensitive to projection effects than depressions in
the X-ray image (as also pointed out by B\^irzan et al. 2008).  By
assuming that the cavities might be as extended as the radio lobes
observed at 235 MHz, we estimate that cavity power could increase by a
factor $\sim 2$ (see Table \ref{cavity.tab}).

Recent studies have shown that the mechanical cavity power usually
exceeds the radio luminosity of the bubbles in galaxy groups and
clusters, with a large scatter from a few to a few thousands that
seems to show a trend with the radio luminosity itself, i.e. the ratio
of cavity power to radio luminosity increases with decreasing radio
luminosities (B\^irzan et al. 2004, 2008).  Thanks to the new {\it
  GMRT} radio data, we can compare the AGN mechanical power with the
radio luminosity of the source in order to directly estimate its
radiative efficiency.  The monochromatic 235 MHz power of the total
radio source is $P_{\rm 235} = 1.8 \times 10^{22} \, {\rm W \,
  Hz}^{-1}$.  By assuming a spectral index $\alpha = 1.2$ (see \S
\ref{radio.sec}), we measure the total radio luminosity over the
frequency range of 10 MHz - 10 GHz to be $3.0 \times 10^{38}$ erg
s$^{-1}$ , which is about four orders of magnitude less than the total
power of the cavities.  Therefore the radio source in HCG 62 has a
synchrotron radiative efficiency as low as $\sim 10^{-4}$. This makes
HCG 62 the least-efficient system in the B\^irzan et al. (2008)
sample, and is a confirmation that the synchrotron radio luminosity is
not a reliable gauge of the total mechanical power of the AGN
outburst, as the radio sources can be very poor or time-variable
radiators.


\begin{deluxetable}{lcccc}[t]
\tablewidth{0pt}
\tablecaption{Cavity properties
\label{cavity.tab}
 }
\tablehead{
\colhead{} & \multicolumn{2}{c}{Cavity N} &  \multicolumn{2}{c}{Cavity S}
\\
\colhead{} & \colhead{(using $V_X$)} &  \colhead{(using $V_{\rm radio}$)} & \colhead{(using $V_X$)} & \colhead{(using $V_{\rm radio}$)} 
}
\startdata
$a$~ (kpc)  & $5$ & $10$ & $4$ & $10$
\\
$b$~ (kpc)  & $4.3$ & $6$ & $4$ & $6$
\\
$R$~ (kpc)  & $8.4$ & $15$ & $8.6$ & $15$
\\
$V \, ({\rm cm}^3)$ & $1.14 \times 10^{67}$ & $4.43 \times 10^{67}$ & $7.88 \times 10^{66}$ & $4.43 \times 10^{67}$
\\
$p \, {\rm (erg \, cm}^{-3})$ & $2.5 \times 10^{-11}$ & $1.8 \times 10^{-11}$ & $3.2 \times 10^{-11}$ &  $2.2 \times 10^{-11}$ 
\\
$p V$~ (erg)& $2.9 \times 10^{56}$ & $8.0 \times 10^{56}$ & $2.5 \times 10^{56}$ & $9.8 \times 10^{56}$
\\
$kT$ (keV) & $0.75$ & $0.84$ & $0.83$ & $0.95$
\\
$c_{\rm s} \, (\rm{km \, s}^{-1}) $ & $440$ & $465$ & $463$ & $495$
\\
$t_{\rm s}$~ (yr) & $1.9 \times 10^7$ & $3.2 \times 10^7$ & $1.8 \times 10^7$ & $3.0 \times 10^7$ 
\\
$P_{\rm cav} \, ({\rm erg \, s}^{-1})$ & $2.0 \times 10^{42}$ & $3.2 \times 10^{42}$ & $1.8 \times 10^{42}$ & $4.2 \times 10^{42}$
\\[-2mm]
\enddata
\tablecomments{Cavity volumes are calculated assuming spherical or
    prolate ellipsoidal shapes with semimajor axis $a$ and semiminor
    axis $b$. In particular, X-ray volumes $V_X$ are estimated from {\it
      Chandra} data and are consistent with the values of B\^irzan et
    al. (2004) and Rafferty et al. (2006), whereas radio volumes $V_{\rm
      radio}$ are estimated from {\it GMRT} observations at 235 MHz (see
    \S \ref{radio.sec}).  Cavity powers $P_{\rm cav}$ are calculated
    assuming $4 pV$ of energy per cavity and the sound crossing
    timescale $t_{\rm s}$.  We adopt the temperature and density 
    values estimated in the NE and SW sectors from {\it
      XMM--Newton} spectra (see top left of Fig. \ref{sectors-proj.fig}).  }
\end{deluxetable}

\subsubsection{Pressure Balance}
\label{pressure.sec}

Since the radio source is filling the cavities, we can compare
directly the radio pressure of the relativistic plasma internal to the
lobes with the X-ray pressure of the surrounding thermal gas.  The
pressure of the hot gas is measured from the density and temperature
derived from the X-ray data as $p = 2 n_{\rm e} k T $ (\S
\ref{deproj.sec}).  In particular, for the X-ray pressure in the
northern and southern cavities we assume the values estimated in the
first bin by performing a deprojection analysis along the NE and SW
sectors, respectively (see top left of
Fig. \ref{sectors-proj.fig}). We find: $p_{\rm X,N} = 2.5 \times
10^{-11} {\rm erg \, cm}^{-3}$, $p_{\rm X,S} = 3.2 \times 10^{-11}
{\rm erg \, cm}^{-3}$.  These values are consistent with the gas
pressure at the radius of the cavities estimated by Morita et
al. (2006).

The total pressure in a radio lobe is the sum of the magnetic
pressure, $p_{B}$, and the total particle pressure, $p_{\rm part}$, and
can be written as
\begin{equation}
  p_{\rm radio} = p_{B} + p_{\rm part} 
= \frac{E_{B}}{\phi V} + \frac{1}{3} \frac{E_{\rm part}}{\phi V} 
= \frac{B^2}{8 \pi} + \frac{1}{3} \frac{(1+k) E_{\rm e}}{\phi V}
\label{pradio.eq}
\end{equation}
where $k$ is the ratio of the energy in protons to that in electrons
($E_{\rm e}$), $V$ is the volume of the radio lobe and $\phi$ is the
volume filling factor.  Using the expression for $E_{\rm e}$ given in
Pacholczyz (1970), Eq. \ref{pradio.eq} determines the lobe pressure in
terms of the magnetic field strength and the factor $k/\phi$, once the
volume $V$ of the radio lobe is known.  This calculation is usually
performed under the widely adopted minimum energy conditions, in which
the relativistic plasma is in equipartition with the magnetic field
($B_{\rm eq}$).  For historical reasons the frequency band adopted to
calculate the standard equipartition field is $\nu_1=10$ MHz -
$\nu_2=100$ GHz, i.e. roughly the frequency range observable with
radio telescopes.  From a physical point of view, the adoption of this
frequency band in the calculation of the minimum energy is equivalent
to the assumption that only electrons emitting between 10 MHz - 100
GHz, i.e. with energy between $\gamma_{\rm min} \propto (\nu_1/B_{\rm
  eq})^{1/2}$ and $\gamma_{\rm max} \propto (\nu_2/B_{\rm eq})^{1/2}$
are present in the radio source.  This approach neglects the
contribution of the electrons emitting below 10 MHz and, as a more
serious bias, in radio sources with different $B_{\rm eq} $ selects
different energy bands of the electron population because the energy
of the electrons which emit synchrotron radiation at a given frequency
depends on the magnetic field intensity (Brunetti 2002).  We therefore
adopt a different approach to calculate the minimum energy conditions,
in which $B_{\rm eq}$ does not depend on the emitted frequency band
but directly on the low energy cut-off of the electron spectrum.
These so-called {\it ``revised''} equipartition conditions select also
the contribution to the energetics due to the low-energy electrons
(Brunetti et al.  1997). In our calculations we assume $\gamma_{\rm
  min} = 100$ and the observed spectral index (see
Tab. \ref{equip.tab}).


\begin{deluxetable}{lcccc}[ht]
\tablewidth{0pt}
\tablecaption{Equipartition calculations
\label{equip.tab}
 }
\tablehead{
\colhead{} & \multicolumn{2}{c}{Cavity N} &  \multicolumn{2}{c}{Cavity S}
\\
\colhead{} & \colhead{\it (revised)} &  \colhead{(standard)} & \colhead{\it (revised)} & \colhead{(standard)} 
}
\startdata
$S_{\rm 235} \, {\rm (mJy)}$ & $6.0$ & $6.0$ & $5.4$ & $5.4$
\\
$S_{\rm 610} \, {\rm (mJy)}$ & $1.8$ & $1.8$ & $1.7$ & $1.7$
\\
$\alpha^{610}_{235}$ & $1.3$ & $1.3$ & $1.2$ & $1.2$
\\
$P_{235} \, (\rm W \, Hz^{-1})$ & $2.5 \times 10^{21}$ & $2.5 \times 10^{21}$ & $2.3 \times 10^{21}$ & $2.3 \times 10^{21}$
\\
$\gamma_{\rm min}$ & $100$ & $900$ & $100$ & $850$
\\
$B_{\rm eq} \, (\mu{\rm G})$ & $6.7$ & $3.1$ & $6.6$ & $3.3$ 
\\
$p_{\rm radio} \, {\rm (erg \, cm}^{-3})$ & $2.3 \times 10^{-12}$ & $4.8 \times 10^{-13}$ & $2.3 \times 10^{-12}$ & $5.5 \times 10^{-13}$
\\
$p_{\rm X}/p_{\rm radio}$ & $10.9$ & $52.1$ & $13.9$ & $58.2$
\\
$k_{\rm bal}$ & $318$ & $1830$ & $500$ & $2450$ 
\\[-2mm]
\enddata
\tablecomments{ Results of equipartition calculations.  We assume
    $\phi=1$, $k=1$ and $V = V_X$ estimated in
    \S \ref{energetics.sec}. A low energy cut-off $\gamma_{\rm min}
    =100$ of the electron spectrum is assumed to calculate the {\it
      ``revised''} equipartition field, which implies a spectral
    frequency range of $\sim$ 280 kHz - 100 GHz.  We also show for
    comparison the results of the standard equipartition calculations,
    which adopt instead a fixed emitted frequency band 10 MHz - 100 GHz.
  }
\end{deluxetable}

The results of the equipartition calculations are reported in Table
\ref{equip.tab}, where for comparison we also show the standard
equipartition values.  We estimate that the X-ray pressure is about
one order of magnitude higher than the radio pressure, as typically
found in cavity systems (e.g., Blanton et al. 2001, De Young 2006,
Croston et al. 2008).  We also find that with revised equipartition
the cavities are closer to pressure balance than they are with
standard equipartition.  On the other hand, by assuming the lobes are
in pressure equilibrium with the ambient gas we can obtain constraints
on the particle content within the radio lobes (Dunn \& Fabian 2004,
B\^irzan et al. 2008).  In particular, we determine the ratio $k_{\rm
  bal}$ of the energy in protons to that in electrons that is required
to achieve pressure balance under revised equipartition conditions.
We find $k_{\rm bal} \sim$ 300-500, whereas higher values ($k_{\rm
  bal} \sim$ few thousands) are expected with standard equipartition.
These values are in the range found by B\^irzan et al. (2008), who
studied the energetics and particle content of the lobes of 24 radio
galaxies in cooling cores, obtaining values of $k_{\rm bal}$ up to
approximately 4000 (with standard equipartition).  Based on an
analysis of nine FR-I radio galaxies in a sample of groups of
galaxies, Croston et al. (2008) found that the radio lobes are
underpressured by a factor up to 70 (with revised equipartition,
assuming $\gamma_{\rm min} = 10$ and $\alpha = 0.5$), and that the
pressure imbalance appears to be linked to the radio-source
morphology, i.e. 'plumed' sources typically have larger pressure
deficits than sources where the jets are embedded in the lobes
('bridged' sources).  The authors interpret this result as evidence
that plumed sources have a higher entrainment rate due to the larger
fraction of the jet surface which is in direct contact with the
external medium, leading to an increase in $k/\phi$.  Although the
classification into bridged and plumed morphologies may not directly
apply to radio sources at the center of cooling core systems,
typically having amorphous structures, this picture is consistent with
the results of Dunn et al. (2006) who argue that the large pressure
imbalance observed in radio bubbles as those of the Perseus cluster is
more likely to be due to entrainment rather than a relativistic proton
population.  Unfortunately we cannot derive good constraints on the
energy in relativistic particles without knowing the break frequency
in the radio spectrum. However, we note that HCG 62 shows a disturbed
radio morphology with inner lobes clearly filling the well defined
X-ray cavities, but with outer lobes having no associated X-ray
cavities (see \S \ref{radio.sec}).  Assuming their detection is not
limited by the sensitivity of the current {\it Chandra} image, this
suggests the possibility of mixing between ambient gas and radio
plasma in the lobes. Therefore the $k_{\rm bal} > 0$ values that we
measure in the lobes might be the results of entrainment of thermal
gas through the group atmosphere.

\subsection{Shock Fronts}
\label{front.sec}

\begin{figure*}[t]
\centerline{
\includegraphics[width=7cm,angle=-90]{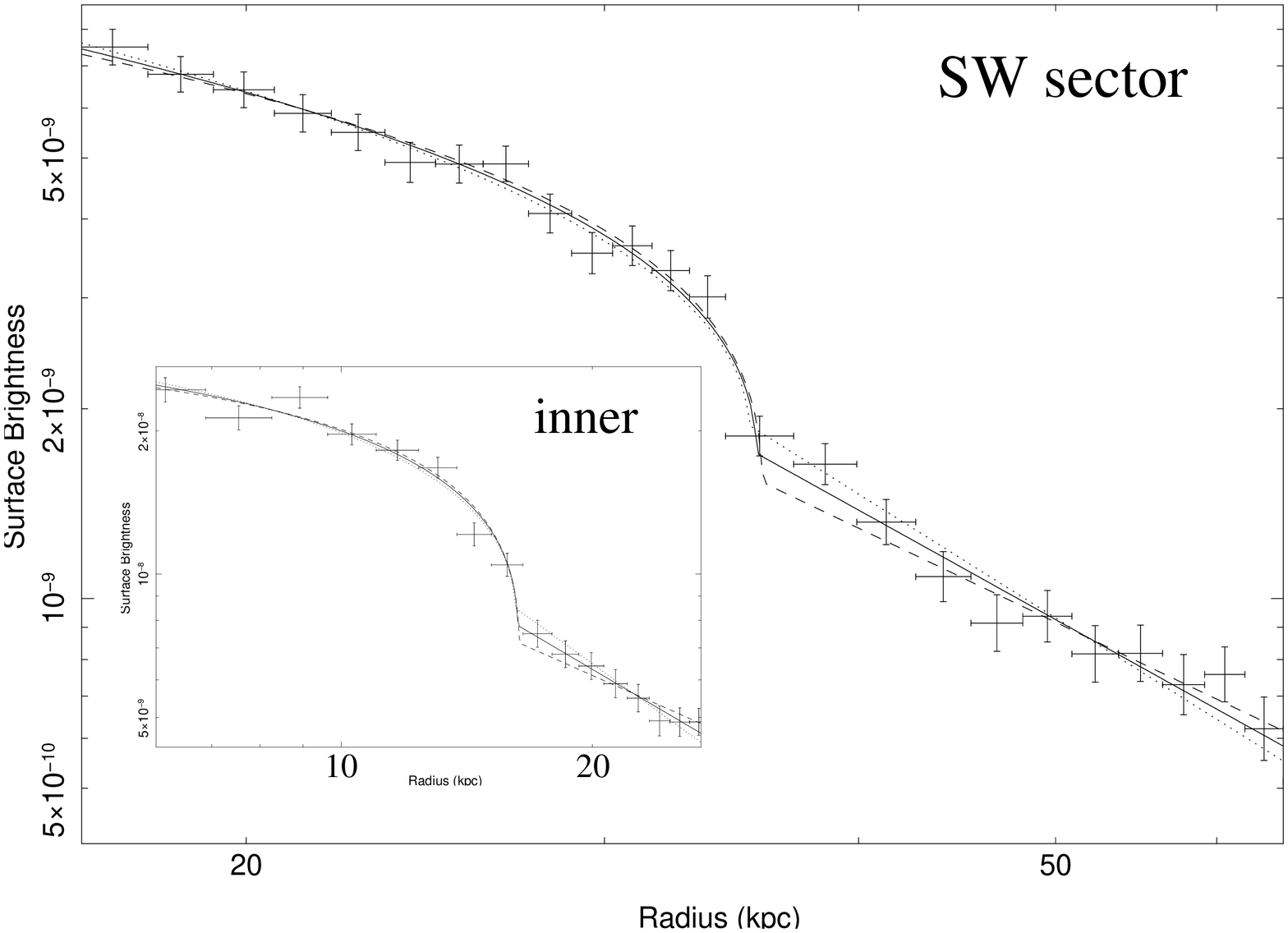}
\includegraphics[width=7cm,angle=-90]{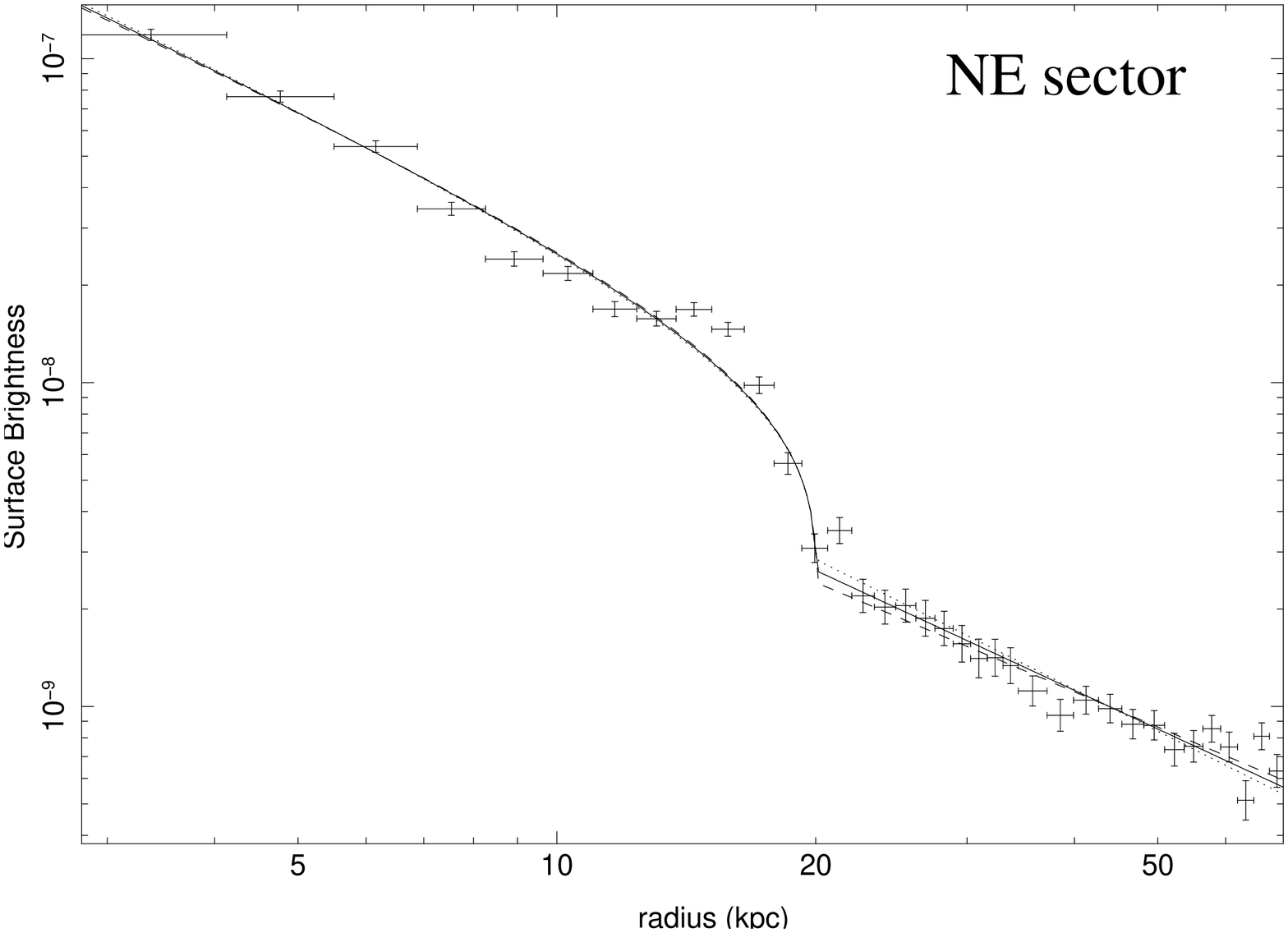}
}
\centerline{
\includegraphics[width=8.5cm]{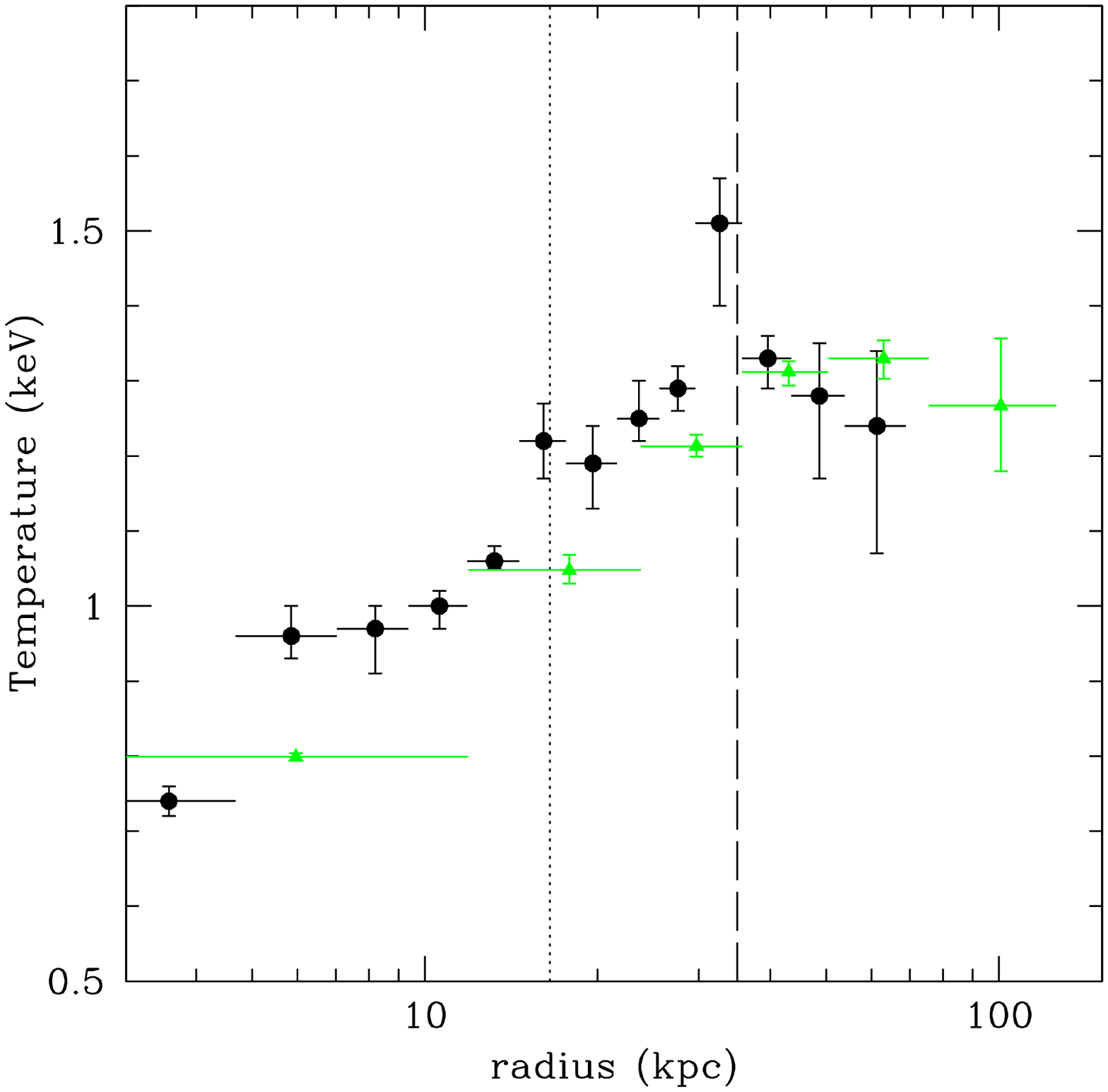}
\includegraphics[width=8.5cm]{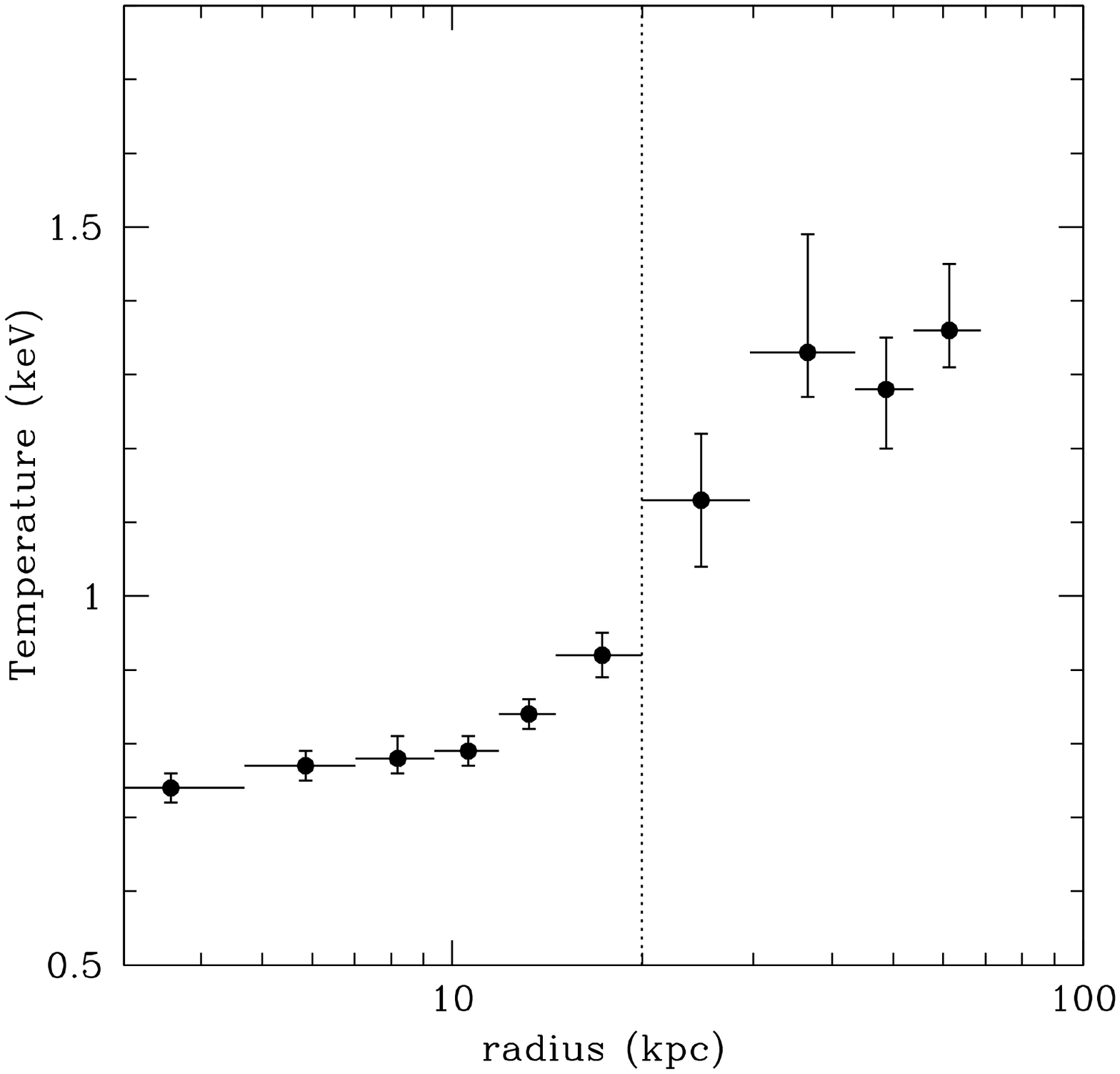}
}
\caption{\label{front.fig} {\it Top, left:} Background-subtracted,
  exposure-corrected {\it Chandra} surface brightness profile
  extracted along the SW sector from P.A. $260^{\circ}$ to
  $330^{\circ}$ in the energy range 0.5-2.0 keV. The surface
  brightness is in counts cm$^{-2}$ s$^{-1}$, with errors at
  1$\sigma$.  Radial error bars show the limits of the bins. The
  smooth curves show fits of the broken power-law density model which
  give density jumps of 1.47 (dotted), 1.65 (solid), and 1.86
  (dashed), corresponding to Mach numbers of 1.32 (dotted), 1.45
  (solid), and 1.62 (dashed). The radius of the shock is 36 kpc.  The
  inner box shows the surface brightness profile of the inner front,
  which is at a radius of 16 kpc.  The solid curve shows the fit of
  the broken power-law density model which gives a density jump of
  1.77.  {\it Top, right:} Similar to top left panel, but for NE
  sector from P.A. $95^{\circ}$ to $160^{\circ}$. The solid curve
  shows the fit of the broken power-law density model which gives a
  density jumps of 2.27. The radius of the front is 20 kpc.  {\it
    Bottom, left:} {\it Chandra} (black circles) and {\it XMM--Newton}
  (green triangles) temperature profiles measured along the SW sector
  (same sector as in top left panel).  The dashed line indicates the
  position of the shock ($r_{\rm shock} = 36$ kpc), while the dotted
  line indicate the position of the inner front ($r_{\rm front,SW} =
  16$ kpc).  {\it Bottom, right:} {\it Chandra} temperature profile
  measured along the the NE sector (same sector as in top right
  panel).  The dotted line indicates the position of the front
  ($r_{\rm front,NE} = 20$ kpc). Error bars on temperature values are
  at 1$\sigma$.  }
\end{figure*}

\begin{figure*}[ht]
\centerline{
\includegraphics[width=8.03cm]{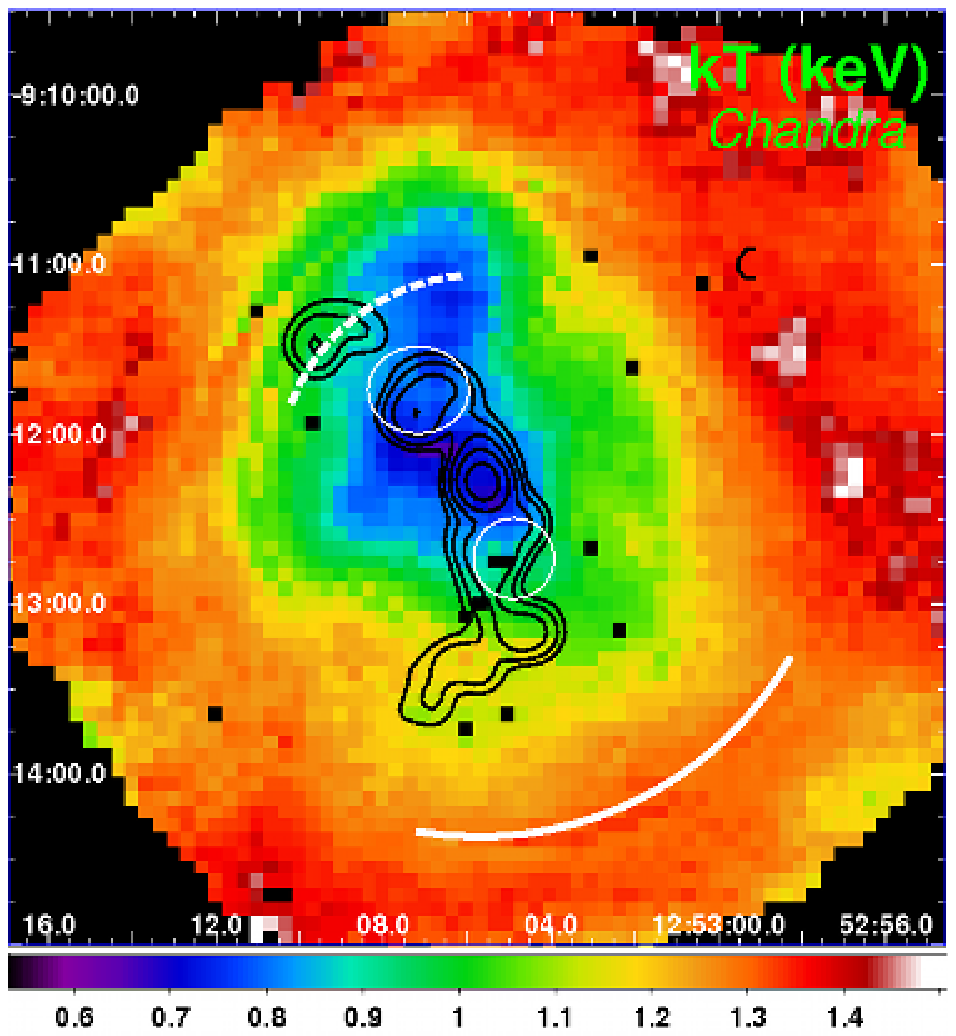}
\includegraphics[width=8.0cm]{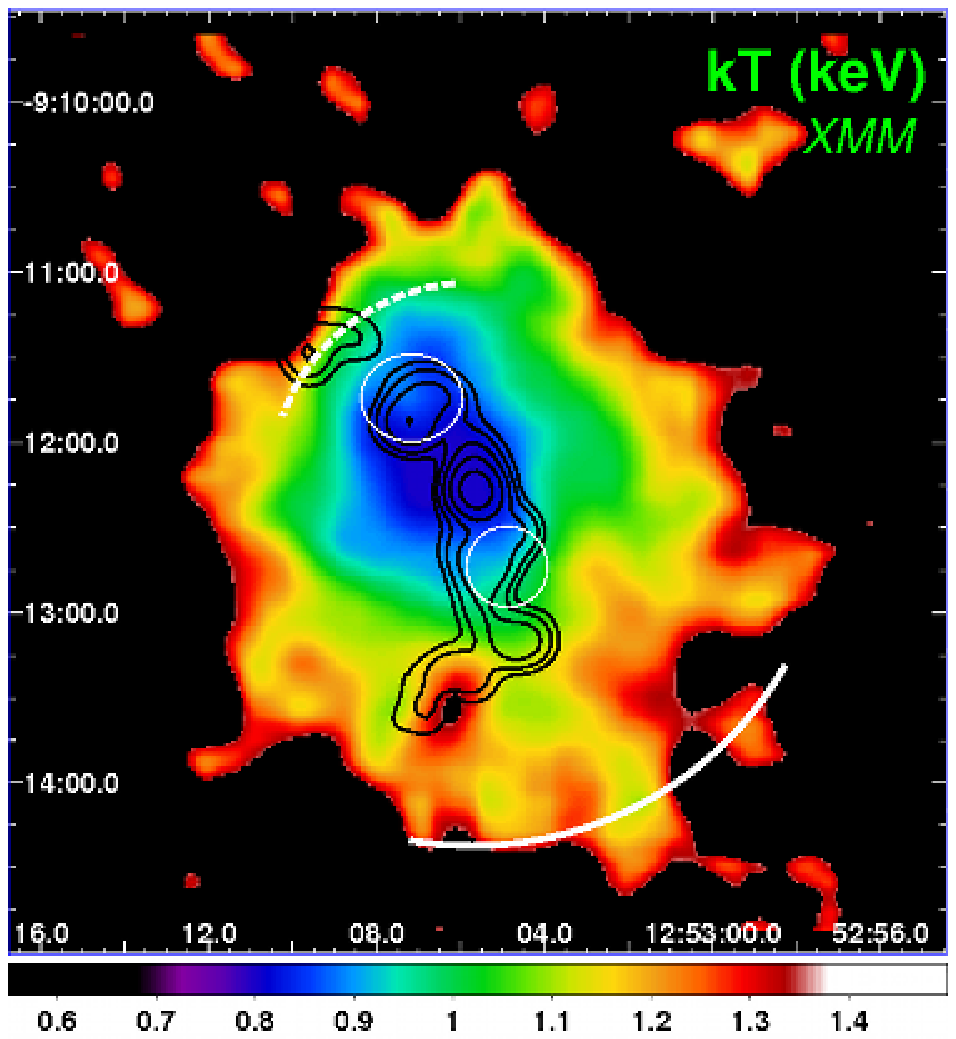}
}
\caption{\label{Tmap.fig} {Left: } {\it Chandra} temperature
  map with GMRT 235 MHz radio emission overlaid in black contours.
  Superposed in white are the two elliptical regions indicating the
  cavities (thin), and two arcs indicating the SW shock front (thick)
  and the NE front (thick dashed).  
  {\it Right: } {\it XMM} temperature map obtained by means of the
  hardness ratio of different energy bands.  Overlaid radio contours
  and superposed regions are the same as in left panel.  Owing to the
  methods adopted to build such maps, the pixel values are not
  independent (see text for details).}
\end{figure*}

We wish to investigate in more detail the morphological features
presented in \S \ref{X-morphology.sec} by measuring the surface
brightness and temperature profiles.

The surface brightness profile extracted along the the SW sector
within an aperture of $260^{\circ}$ - $330^{\circ}$ (counting
counterclockwise from P.A. = $0^{\circ}$ toward the west) is shown in
the top-left of Fig. \ref{front.fig}.  By fitting a broken power-law
density model, at 36 kpc from the center we identify a clear break in
surface brightess which corresponds to a jump in density of 1.65.  To
unambiguously determine the nature of this front, we extracted the
temperature profile along the same sector (Fig. \ref{front.fig},
bottom left) and find that the region immediately interior to the
front is significantly hotter than the undisturbed region just outside
of it, with a temperature jump across the front of $\sim$14\%.  As
discussed below, this is consistent with the interpretation of the
front as a shock.

We also identify a front to the NE at 20 kpc from the center (see Fig.
\ref{front.fig}, top right), that appears to have a symmetric feature
to the SW (the surface brightness profile shows an edge at 16 kpc from
the center, see inner box in Fig. \ref{front.fig}, top left). We note
that the detection of a temperature rise in the regions immediately
inside of these fronts, which would be expected if they are shocks, is
complicated by the underlying rising temperature profile of the global
group atmosphere (see Fig. \ref{front.fig}, bottom).  For this reason,
despite the lack of evidence for a temperature jump across the inner
fronts, we cannot rule out the presence of shocks. In fact, the
symmetric placement of these fronts and their proximity to the
cavities (3-5 kpc outside the cavity edge) support their
interpretation as weak shocks.  Alternatively, they could be
interpreted as cold fronts, caused by the subsonic expansion of the
cavities displacing cool, low entropy gas outward.  However, since we
cannot place conclusive constraints on these features with the present
data, we will not discuss them further in the paper and in the
following we will focus only on the outer shock detected to the SW
(simply referred as to 'the shock').

A hydrodynamic model for the shock was made by initiating an explosion
at the center of a hydrostatic, isothermal atmosphere with a power-law
density profile. The power-law index for the density for the unshocked
gas was determined from the broken power-law fit to the surface
brightness profile.  Our data are consistent with the presence of a
shock having a Mach number $\mathcal{M} = 1.45$ and an age \gtsim
$2.7 \times 10^7$ yr.  Compared to a model with continuous energy
injection, the point explosion produces a stronger, i.e. faster, shock
at early times so its age estimate is a lower limit.  For such a Mach
number our model predicts the emission measure weighted temperature to
rise by $\sim$15\% across the front, which is in agreement with our
measurements.  As it appears evident from the comparison with the
temperature profile obtained with {\it XMM--Newton} along the same
sector (overlaid in green in Fig. \ref{front.fig}, bottom left), the
detection of the temperature jump is possible only with the superb
spatial resolution of {\it Chandra}. We estimate an energy $\sim 3.4
\times 10^{58}$ erg, and a power $\sim 4.0 \times 10^{43} \, {\rm erg
  \, s}^{-1}$ which is about one order of magnitude higher than the
cavity power.  The unusually large difference between the shock power
and the cavity power is essentially due to the fact that the shock is
much further away than the cavities from the group center ($R_{\rm
  shock} = 36$ kpc vs. $R_{\rm cav} \sim 8.5$ kpc). Such a difference
may be reduced by a factor $\sim 2$ if we adopt the cavity volume
estimated from the radio data instead of that estimated from X-rays.

From a physical point of view, the difference in location and
energetics between the shock and cavity system can be explained by two
possibilities: 1) the shock and cavities are created by the same,
violent AGN outburst; 2) the shock and cavities originate from
different, multiple episodes of gentle AGN outbursts.
1) In the first case, the jet is initially supersonic and inflates the
cavity violently, with much of the energy driving a shock.  Once the
jet slows the cavity rises buoyantly and the shock detaches from its
tip and continues to propagate in the group atmosphere at a velocity
$\mathcal{M} c_s$.  By assuming that the shock was never any weaker
than it is now and that the group atmosphere has an average
temperature of $\sim$1 keV, we estimate the time for the shock to
travel from the AGN to its current location to be \ltsim $4.8 \times
10^7$ yr, which is consistent with the shock age estimated above.  The
shock age is also consistent with the refill timescale of the
cavities, but slightly longer than the buoyancy timescale (see \S
\ref{energetics.sec}). The buoyancy timescale can be considered as an
upper limit to the age of the outburst that produced the cavities,
since it assumes the cavities were inflated at the group core and have
risen slowly. If they formed close to their current location their age
would be even shorter, and the conflict with the shock age more
serious.
2) In the second case, it is possible that the shock originated during
a previous AGN outburst which produced the outer radio lobes and
perhaps an associated pair of cavities, though these are not detected
in the existing data.  In this scenario, the inner cavities are
thought to be evacuated by the inner radio lobes during a more recent
outburst, thus providing a natural explanation for the marginal
difference between the buoyancy timescale of the cavities and the age
of the shock. The expected older outer cavities might be detected with
deeper exposures in a region close to the observed shock front.

However, with the current radio data we are not able to determine the
relationship of the outer radio lobes to the inner ones.  Therefore,
considering also the uncertainties in the timescales estimated above,
we cannot determine whether the observed radio/X-ray morphology is the
result of one single AGN outburst, or multiple episodes.  In either
case, the position of the shock outside the S radio lobe makes
plausible the interpretation of the shock as being directly driven by
the lobe expansion triggered by an AGN outburst.  Averaging over the
past few 10$^7$~yr, the cavity power alone thus provides a lower limit
to the true total mechanical power of the AGN.  Inclusion of the
energy in the shock provides a more complete estimate.  The shock plus
cavity power of $\sim 4.4 \times 10^{43} \, {\rm erg \, s}^{-1}$ is
much higher than the cooling luminosity ($L_X = 1.5 \times 10^{42}
{\rm \, erg \, s}^{-1}$, estimated in \S \ref{energetics.sec}).  If
the power level of $4.4 \times 10^{43} \, {\rm erg \, s}^{-1}$ is
sustained, since it exceeds the power radiated by cooling core, either
the core is currently being heated, or most of the power must be
deposited in gas beyond the cooling core.  The fact that the shock
front is located at the outer edge of the cooling region is consistent
with the latter.

On the other hand, the apparent absence of an outer shock front to the
NE direction, which would be expected if the shock was created by the
two outer radio lobes, requires some asymmetry.  It is also possible
that the shock strength varies with the position angle. In this case
the shock front may be in fact intrinsically symmetric and surrounding
the whole central region of the group, but is detected only to the SW
because of some difference in gas properties (e.g., higher density).
We investigate further these possibilities also by means of a
temperature map, which shows a lack of very cool ($<$0.9 keV) material
on the SW side of the core (see \S \ref{tmap.sec}) that might be
explained by the shock.

\subsection{Spectral Maps}
\label{maps.sec}

Since no obvious feature associated with the cavities and radio lobes
appear in the wedge maps (\S \ref{azimuthal.sec}), we produced
spectral maps with higher spatial resolution in order to investigate
the 2-D temperature (\S \ref{tmap.sec}) and iron abundance (\S
\ref{Zmap.sec}) distribution independently of the choice of the
sectors used to extract the spectra. Such spectral maps are suitable
for identifying potentially interesting structures.

{\it Chandra} spectral maps were produced using a method similar to
that described in O'Sullivan et al. (2005). The ACIS-S3 field-of-view
was divided into a grid of square map pixels with side length 10
physical pixels ($\sim$5$^{\prime\prime}$). For each of these map
pixels a spectrum was extracted from a circular region centered on the
pixel, with a radius chosen so as to obtain 800 net counts in the
0.3-5.0 keV band. Radii were allowed to vary between 10 and 100
physical pixels.  Spectra and responses were created in an identical
fashion to that used in our main spectral analysis and fitted with an
absorbed {\ttfamily apec} model. Point sources were excluded from both
spectra and the calculation of the spectral extraction region size and
energies outside the 0.3-5.0 keV band were excluded.  90\%
uncertainties on parameters were estimated and any pixel with a
temperature uncertainty greater than 15\% was excluded.  The colors
of the remaining map pixels indicate the best fitting temperature, or
abundance.

\subsubsection{Temperature}
\label{tmap.sec}

In Fig. \ref{Tmap.fig} (left) we show the temperature map obtained
from {\it Chandra} spectra.  We also show in Fig. \ref{Tmap.fig}
(right) a temperature image of the group central region built from
{\it XMM--Newton} colors.  Specifically, we extracted mosaiced MOS
images in four different energy bands (0.5-1.0 keV, 1.0-2.0 keV,
2.0-4.5 keV and 4.5-8.0 keV), subtract the background and divide the
resulting images by the exposure maps.  A temperature in each square
map pixel, chosen to have side length of 5 physical pixels
($=5.5^{\prime\prime}$), is then obtained by fitting a thermal plasma
model to the count rate estimated in the above energy bands in a
surrounding area with a fixed radius of 15 physical pixels
($=16.5^{\prime\prime}$).

The two temperature maps are consistent with each other and with
previous results (Morita et al. 2006; Sanders et al. 2009), showing
the presence of the cool core, but also the presence of cool gas in
the region of the northern cavity. Owing to emission weighting in the
large map extraction regions, this cold feature could be related to
the presence of the cavity rims, which are indeed typically observed
to be cold (McNamara \& Nulsen 2007, and references therein).  On the
other hand, the lack of a comparable cold feature in the region of the
southern cavity is notable.  One possible explanation is that there
was initially the presence of cold gas also to the south, but that
this has then been heated by the passage of the shock. This argument
could be taken as an indication of an asymmetric shock, or at least of
an asymmetric shock strength, as discussed in \S \ref{front.sec}.

\subsubsection{Iron abundance}
\label{Zmap.sec}

\begin{figure}[t]
\centerline{
\includegraphics[width=8cm]{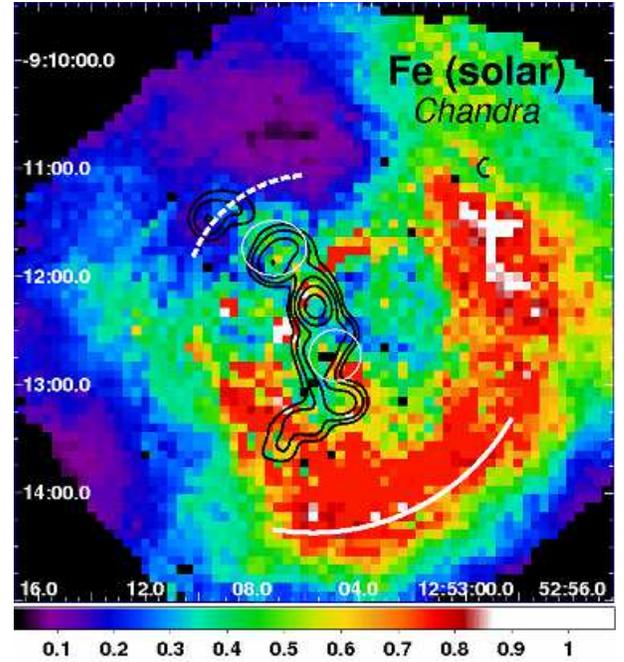}
}
\caption{\label{Zmap.fig} {\it Chandra} iron abundance map
  with GMRT 235 MHz radio emission overlaid in black contours.
  Superposed in white are the two elliptical regions indicating the
  cavities (thin), and the two arcs indicating the SW shock front
  (thick) and the NE front (thick dashed).}
\end{figure}

In Fig. \ref{Zmap.fig} we show the iron abundance map obtained from
{\it Chandra} spectra.  The iron abundance distribution appears to be
inhomogeneous, asymmetric, and lacking the usual central peak.  We
have reason to believe that the structures seen in the abundance map
are real, as they are also observed in the maps of 90\% upper and
lower bound on abundance (not shown here). However, the uncertainties
on the value of individual map pixels is large (30-50\%) and the
variable sizing of spectral extraction regions produces an effect
analogous to adaptive smoothing. It is therefore necessary to extract
separate spectra to determine the true abundance of individual regions
and the significance of any abundance differences.  In particular, to
confirm the presence of the arc-like region of enriched material
visible at about 2$'$ to the SW of the group center, first discovered
by Gu et al. (2007), we extract spectra from two elliptical regions
centered at $2'$ to the NE and SW of the group center along the
direction of the inner radio lobes.  The semimajor axis of both
ellipses is orthogonal to the direction of the inner radio lobes and
$1'$ in extent, with the semiminor axis being half of it. We measure
abundances of $0.12^{+0.06}_{-0.04}$ solar and $0.95^{+0.26}_{-0.16}$
solar (1$\sigma$ errors) in the NE and SW regions, respectively. This
indicates that the SW arc is more abundant than the NE dip at
5$\sigma$ significance.

Gu et al. (2007) discuss several possibilities about the origin of the
high-abundance gas, including entrainment during the cavity expansion
triggered by the AGN activity, and ram pressure stripping due to tidal
interaction of two central galaxies which have experienced a recent
($\sim 10^8$ yr) minor merger, as indicated by optical studies
(Spavone et al. 2006).  Unfortunately, we do not have enough
information in the current data to disentangle this problem and the
origin of the high-abundance arc-like region is still an open
question.
However, we stress here that the enriched material has a location and
shape consistent with that of the newly discovered shock.  Recent
papers by Kaastra et al. (2009) and Prokhorov (2009) claim that in
some cases, for instance near interfaces of hot and cold gas and near
shocks, the usual approximation of Maxwellian electron distributions
adopted for calculating thermal X-ray spectra is no longer valid. In
such situations, the presence of non-thermal electrons modifies the
line emissivities hence the X-ray spectrum, thus affecting the
measurements of metal abundances. In particular, the best-fit iron
abundance for the isothermal model is about 30\% higher than the
actual abundance (Kaastra et al. 2009).  This could explain the high
abundance arc that we measure. A similar effect of apparent iron
overabundance may also be produced by the presence of non-equilibrium
ionization states below the shock (Prokhorov 2010).  These
interpretations of the arc-like feature observed in the abundance map
of HCG 62 are consistent with the presence of the detected shock
front.

\section{Summary}
\label{summary.sec}

We have analyzed new low-frequency {\it GMRT} radio data and existing
{\it Chandra} and {\it XMM--Newton} X-ray data of the compact group
HCG 62 in order to study the properties of the cavity system and the
interplay with the central radio source.  Our investigation
demonstrates the power of a combined X-ray/radio approach to the
feedback problem, and particularly the benefits of extending radio
studies of AGNs to low frequencies where less energetic, older
electron populations are visible.  We summarize our main results
below.

\begin{enumerate}
\item 
In contrast with the early classification of the HCG 62 cavity
  system as ``radio ghost'', the new {\it GMRT} observations detect
  low-frequency radio emission in the cavities.
\item
 The radio source has a total radio luminosity of $3.0 \times
  10^{38}$ erg s$^{-1}$ and a very low radiative synchrotron
  efficiency of $\sim 10^{-4}$.
\item 
  We estimate that the X-ray pressure of the cavities is about one
  order of magnitude higher than the ``revised'' equipartition
  pressure of the radio lobes. By assuming pressure balance, we find
  that the ratio of the energy in protons to that in electrons is
  $\sim$300-500, which may originate from thermal gas entrainment in
  the lobes through the group atmosphere.
\item
 We discover a Mach $\sim$1.5 shock front at 36 kpc from
  the center toward the SW, which has not been reported in
  literature.  The total energy in cavities and shock is $\sim 3.6
  \times 10^{58}$ erg.
\item 
  We confirm the presence of an arc-like region of enriched gas to the
  SW whose apparent high metallicity could be due to a non-Maxwellian
  electron distribution or non-equilibrium ionization states near the
  shock front.
\end{enumerate}

\acknowledgments 

We thank the anonymous referee for constructive suggestions that
improved the presentation of the results.  We thank S. Ettori for
providing the software required to produce the {\it XMM--Newton}
color map in Fig. \ref{Tmap.fig} (right) and M. Markevitch for useful
discussions. M. Gitti also wishes to thank T. Ponman and F. Brighenti
for comments on the original manuscript.  M. Gitti acknowledges
support by grant ASI-INAF I/088/06/0.  E. O'Sullivan acknowledges the
support of the European Community under the Marie Curie Research
Training Network. This research is supported by Chandra grant
GO0-11003X.

\end{document}